\documentclass[1p,times,preprint]{elsarticle}
\usepackage[english]{babel}
\usepackage{tensor}
\usepackage{graphicx}
\usepackage{amsmath}
\usepackage{amssymb}
\usepackage{amsfonts}
\usepackage{dcolumn}
\usepackage{bm}
\usepackage{xcolor}
\usepackage{ulem}
\usepackage{tikz}
\usepackage{subcaption}
\usepackage{comment}
\usepackage{verbatim}
\usepackage{fancyvrb}
\usepackage{cancel}
\usepackage{multirow}
\usepackage{lscape}
\usepackage{txfonts}
\usepackage{mathtools}
\usepackage{soul}
\usepackage{url}
\usepackage{longtable}
\usepackage{makecell}
\usepackage[pdftex]{pict2e}
\usepackage{microtype}

\def\bra<#1|{\mathinner{\langle\,{#1}\,\vert}} 
\def\ket|#1>{\mathinner{\vert\,{#1}\,\rangle}} 
\def\red|#1|{\mathinner{\!\vert\,{#1}\,\vert\!}}
\def\braket<#1>{\mathinner{\langle\,{#1}\,\rangle}} 

\def\redmem#1#2#3{  \left\langle #1 \left\Vert  
                  #2 \right\Vert #3 \right\rangle   }

\begin{document}

\title{ SECOND-ORDER RAYLEIGH-SCHR\"ODINGER PERTURBATION THEORY FOR THE GRASP2018 PACKAGE:
THREE-PARTICLE FEYNMAN DIAGRAM CONTRIBUTION TO VALENCE-VALENCE CORRELATIONS}
\date{\today}

\author[TFAI]{G. Gaigalas}
\address[TFAI]{Institute of Theoretical Physics and Astronomy, Faculty of Physics,
               Vilnius University, Saul\.{e}tekio Ave. 3, LT-10257 Vilnius, Lithuania}
\ead{gediminas.gaigalas@tfai.vu.lt}

\author[TFAI]{P. Rynkun}
\ead{pavel.rynkun@tfai.vu.lt}

\author[TFAI]{L. Kitovien\.{e}}
\ead{laima.radziute@tfai.vu.lt}

%
%
\begin{abstract}
The method based on the second-order perturbation theory 
to identify the most important configuration state functions of the various correlations 
is extended to include valence-valence correlations, which are described 
by the three-particle Feynman diagram. The extension presented 
in this work complements the core-valence, core, core-core and valence-valence correlations 
which were developed in a series of previous papers by G. Gaigalas, P. Rynkun and L. Kitovienė. 
Whereas these valence-valence correlations are described by the three-particle Feynman diagram, 
additional developments to calculate the spin-angular parts of this diagram have been made 
to the program library \texttt{librang} of the {\sc Grasp}. As an example of the application of the developed method, 
the atomic calculations of the energy structure for the Se~III are presented. In the present work, 
this method was also used to select the most significant configuration state functions 
and to use this basis to solve the self-consistent field equations.
\end{abstract}

\begin{keyword}
configuration interaction \sep spin-angular integration \sep perturbation theory \sep tensorial algebra \sep valence-valence  correlations \sep core-valence correlations \sep core correlations \sep core-core correlations


\end{keyword}
\maketitle


\section{Introduction}

An approach based on a combination of the relativistic configuration interaction method and the stationary second-order Rayleigh-Schr\"odinger many-body perturbation theory (RSMBPT) in an irreducible tensorial form has recently been developed in Refs.~\cite{Gaigetal:2024CV,Gaigetal:2024C,Gaigetal:2024CC,Gaigetal:2025VV}.
It allows us to analyze effectively core-valence (CV), core (C), core-core (CC) and valence-valence (VV) correlations using the General Relativistic Atomic Structure package {\sc Grasp}~\cite{grasp2018}, which are described by vacuum, one- and two-particle Feynman diagrams. At the same time, this theory extends the potential of the  methodology~\cite{Fisetal:16a,Jonsetal:23} used worldwide.

The formulation of this theory is based on the second-order Linked-Diagram Theorem \cite[Section 12.5.2]{LindgrenBook:82}, where the lines of the Feynman diagrams corresponding to the electronic excitations are linked together to form a solid line. Based on this theorem, we further develop the capabilities of the theory in this paper. For completeness, we include in the paper (Section~\ref{sec:seconOrder}) the three-particle Feynman diagram describing the valence-valence correlations. Since the spin-angular part of this diagram is much more complex, a major part of the paper is devoted to calculation of the spin-angular coefficients (Section~\ref{Sec:spin_angular}). The final expressions are presented in Section~\ref{Sec:VVImplementation} of the paper. Section~\ref{Sec:Calculations} contains calculations showing that the expressions and the examination of the valence-valence correlations presented in the paper are correct and usable. Conclusions are presented in Section~\ref{Sec:Conclusions}.

\section{Relativistic second-order effective Hamiltonian of an atom or ion in irreducible tensorial form for the remaining valence-valence correlations}
\label{sec:seconOrder}

New Feynman diagrams are used to describe valence-valence correlations, which were not available for core-valence, core, and core-core correlations. We discuss this in detail below.

\subsection{The third type of valence-valence correlations}
\label{sec:PT_Mano_Third}

The first two types of valence-valence correlations are already described in the paper~\cite{Gaigetal:2025VV}.
The third type of valence-valence correlations is presented through the Feynman diagram VV$_3$ from Fig.~\ref{VV_3} where all lines with the double arrow of diagram are renamed in the following way: $m'=p \equiv m$ and $m=n=n'=p' \equiv n$.
\begin{equation}
\label{eq:VVT-a} 
(n_{m} \ell_{m}) \; j_{m}^{w_m} \; (n_{n} \ell_{n}) \; j_{n}^{w_n}
   \rightarrow (n_{m} \ell_{m})\, j_{m}^{w_m+1} \; (n_{n} \ell_{n}) \; j_{n}^{w_n-2}
	 \; (n_{r} \ell_{r}) \; j_{r}.
\end{equation}

\begin{figure*}
\begin{center}
\setlength{\unitlength}{1mm}
\begin{picture}(180,43)(4,0)
\thicklines
\put(10,30){\line(0,1){10}}
\put(10,38){\vector(0,1){2}}
\put(10,40){\vector(0,1){2}}
\put(7.5,38){\makebox(0,0)[t]{\small{$m$}}}
\put(17.5,38){\makebox(0,0)[t]{\small{$n$}}}
\multiput(10,30)(1,0){10}{\circle*{0.35}}
\put(10,20){\line(0,10){10}}
\put(17.6,24){\makebox(0,0)[r]{\small{$r$}}}
\put(20,38){\vector(0,1){2}}
\put(20,40){\vector(0,1){2}}
\put(20,24){\vector(0,1){3}}
\put(20,10){\line(0,10){30}}
\put(20,10){\vector(0,1){3}}
\put(20,08){\vector(0,1){3}}
\put(17.5,13){\makebox(0,0){\small{$n^{\prime}$}}}
\multiput(20,20)(1,0){10}{\circle*{0.35}}
\put(10,10){\line(0,1){10}}
\put(10,10){\vector(0,1){3}}
\put(7.5,13){\makebox(0,0){\small{$m^{\prime}$}}}
\put(10,08){\vector(0,1){3}}
\put(30,10){\line(0,1){20}}
\put(30,30){\vector(0,1){3}}
\put(30,28){\vector(0,1){3}}
\put(27.5,28){\makebox(0,0){\small{$p$}}}
\put(30,10){\vector(0,1){3}}
\put(30,08){\vector(0,1){3}}
\put(27.5,13){\makebox(0,0){\small{$p^{\prime}$}}}
\put(20,04){\makebox(0,0){$\text{VV}_{3}$}}
\put(37,37){\makebox(0,0) [l] {$\displaystyle{ = -
	\sum_{m, m^{\prime}}~\sum_{n, n^{\prime}}~\sum_{p, p^{\prime}}~\sum_{k, k^{\prime}, x}~~\left( -1 \right)^{j_{n} + j_{n^{\prime}} + k + k^{\prime}} \sqrt{\frac{\left[ x \right]}{\left[ k, k^{\prime} \right]}}
	}$}}
\put(40,25){\makebox(0,0) [l] {$\displaystyle{ \times
		\left[\left[\left[\;  a^{\left( j_m \right) }  \times 
  \tilde a^{\left( j_{m^{\prime}} \right) } \; \right] ^{\left( k \right)} \times 
	\left[\; a^{\left( j_{n} \right) }  \times 
  \tilde a^{\left( j_{n^{\prime}} \right) } \; \right] ^{\left( x \right)} \right]^{\left( k^{\prime} \right)}	
  \times
	\left[\; a^{\left( j_{p} \right) }  \times 
  \tilde a^{\left( j_{p^{\prime}} \right) } \; \right] ^{\left( k^{\prime} \right)} \right]^{\left( 0 \right)}
	}$}}
\put(40,13){\makebox(0,0) [l] {$\displaystyle{ \times
\sum_{r}~\frac{1}
{\left( \varepsilon_{n'}+\varepsilon_{p'}-\varepsilon_r-\varepsilon_p \right)}
	\left\{
    \begin{array}{ccc}
      j_{n'} & j_{n} & x \\
      k      & k'    & j_{r}  
     \end{array}
	\right\}  
	X_{k}(m n, m' r) ~ X_{k'}(r p, n^{\prime} p^{\prime})}$}}
\end{picture}
\caption{The VV Feynman diagram of the second-order effective Hamiltonian for the third and fourth types of valence-valence correlations 
$(n_{m} \ell_{m})\, j_{m}^{w_m} \, (n_{n} \ell_{n})\, j_{n}^{w_n} 
   \rightarrow (n_{m} \ell_{m})\, j_{m}^{w_m+1} \; (n_{n} \ell_{n})\, j_{n}^{w_n-2} \; (n_{r} \ell_{r})\, j_{r}$ 
and
$(n_{m} \ell_{m})\, j_{m}^{w_m} \, (n_{n} \ell_{n})\, j_{n}^{w_n} \, (n_{p} \ell_{p})\, j_{p}^{w_p} 
   \rightarrow (n_{m} \ell_{m})\, j_{m}^{w_m+1} \; (n_{n} \ell_{n})\, j_{n}^{w_n-1} \; (n_{p} \ell_{p})\, j_{p}^{w_p-1} \; (n_{r} \ell_{r})\, j_{r}$.
}
\label{VV_3}
\end{center}
\end{figure*}

The three-particle Feynman diagram's VV$_3$ expression, like the diagrams described in the previous papers~\cite{Gaigetal:2024CV,Gaigetal:2024C,Gaigetal:2024CC,Gaigetal:2025VV}, have an energy denominator
$D = \sum \left( \varepsilon_{\text{down}} - \varepsilon_{\text{up}} \right)$,
where $\varepsilon_{\text{down}}$ ($\varepsilon_{\text{up}}$) is the single-particle eigenvalue associated with the down- (up-)
orbital lines to (from) the lowest interaction line of the diagram. For example, the denominators for the VV$_3$ diagram are
\begin{equation}
\label{eq:denominator}
D = \left( \varepsilon_{n'}+\varepsilon_{p'}-\varepsilon_r-\varepsilon_p \right) ,
\end{equation}
where indexes $n'$, $p'$ and $p$ belong to $F'$ set, and $r$ belong to $G$ set of orbitals~\cite{Gaigetal:2024CV}. 

Also, the following notations are used in the expressions of this diagram (see Fig.~\ref{VV_3}):
\begin{equation}
\label{eq:deffX}
   X_{k}(i j, i' j') 
   = \redmem{\ell_i j_{i}}{\, C^{(k)} \,}{ \ell_{i'} j_{i'}}
     \redmem{\ell_j j_{j}}{\, C^{(k)} \,}{ \ell_{j'} j_{j'}} 
R^{k}(n_i j_i \, n_jj_j, \, n_{i'}j_{i'} \, n_{j'}j_{j'} ) ,
\end{equation}
where $R^{k}\left(n_i j_i \, n_jj_j, \, n_{i'}j_{i'} \, n_{j'}j_{j'} \right)$ is the radial integral 
of electrostatic interaction between electrons~\cite[(89) and (90)]{Fisetal:16a} and 
$\redmem{\ell_i j_{i}}{\, C^{(k)} \,}{ \ell_{i'} j_{i'}}$ is the reduced matrix element of the irreducible tensor operator $C^{(k)}$ in $jj$-coupling.

In contrast to the previous diagrams (see Refs.~\cite{Gaigetal:2024CV,Gaigetal:2024C,Gaigetal:2024CC,Gaigetal:2025VV}), this diagram has six operators of second quantization (three pairs of creation and annihilation operators). This complicates the finding values of the spin-angular coefficient of this diagram and therefore makes the program library \texttt{librang}~\cite{Gaigalas:2022} from {\sc Grasp} unusable in the most general case. 
This problem will be discussed and solved in detail in the next section.

\subsection{The fourth type of valence-valence correlations}
\label{sec:PT_Mano_Fourth}

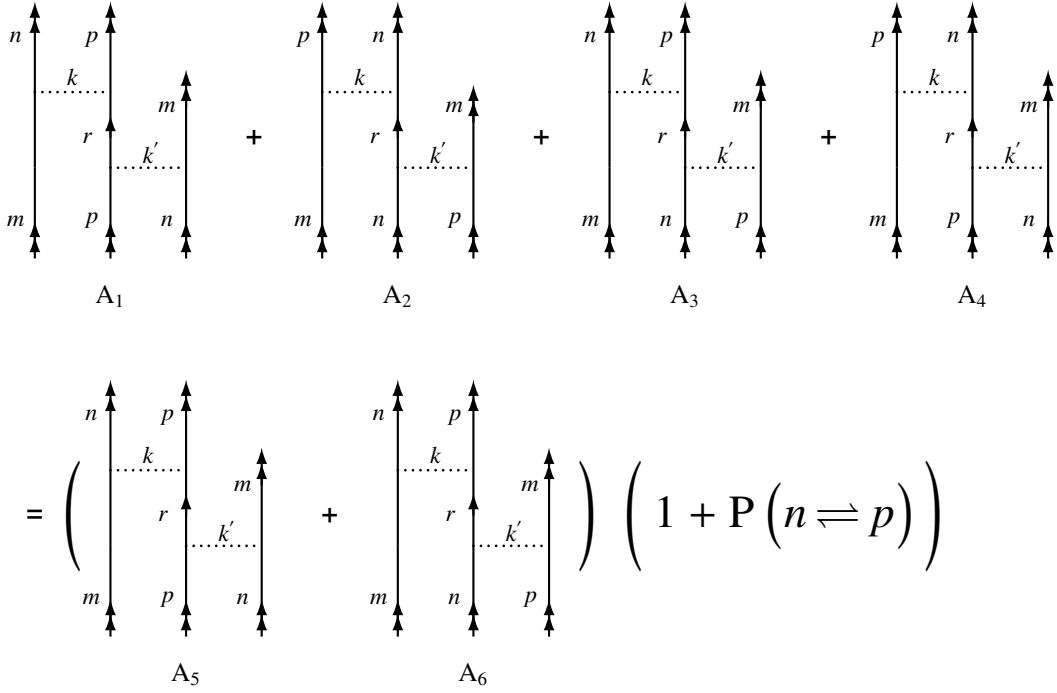
\begin{figure*}
\begin{center}
\setlength{\unitlength}{1mm}
\begin{picture}(180,93)(8,0)
\thicklines
\put(10,80){\line(0,1){10}}
\put(10,88){\vector(0,1){2}}
\put(10,90){\vector(0,1){2}}
\put(7.5,88){\makebox(0,0)[t]{\small{$n$}}}
\put(17.5,88){\makebox(0,0)[t]{\small{$p$}}}
\put(15,82){\makebox(0,0){\small{$k$}}}
\multiput(10,80)(1,0){10}{\circle*{0.35}}
\put(10,70){\line(0,10){10}}
\put(17.6,74){\makebox(0,0)[r]{\small{$r$}}}
\put(20,88){\vector(0,1){2}}
\put(20,90){\vector(0,1){2}}
\put(20,74){\vector(0,1){3}}
\put(20,60){\line(0,10){30}}
\put(20,60){\vector(0,1){3}}
\put(20,58){\vector(0,1){3}}
\put(17.5,63){\makebox(0,0){\small{$p$}}}
\put(25.5,72){\makebox(0,0){\small{$k^{'}$}}}
\multiput(20,70)(1,0){10}{\circle*{0.35}}
\put(10,60){\line(0,1){10}}
\put(10,60){\vector(0,1){3}}
\put(7.5,63){\makebox(0,0){\small{$m$}}}
\put(10,58){\vector(0,1){3}}
\put(30,60){\line(0,1){20}}
\put(30,80){\vector(0,1){3}}
\put(30,78){\vector(0,1){3}}
\put(27.5,78){\makebox(0,0){\small{$m$}}}
\put(30,60){\vector(0,1){3}}
\put(30,58){\vector(0,1){3}}
\put(27.5,63){\makebox(0,0){\small{$n$}}}
\put(20,53){\makebox(0,0){$\text{A}_{1}$}}
\put(39,74){\makebox(0,0){\normalsize{\bf{+}}}}
\put(48,80){\line(0,1){10}}
\put(48,88){\vector(0,1){2}}
\put(48,90){\vector(0,1){2}}
\put(45.5,88){\makebox(0,0)[t]{\small{$p$}}}
\put(55.5,88){\makebox(0,0)[t]{\small{$n$}}}
\put(53,82){\makebox(0,0){\small{$k$}}}
\multiput(48,80)(1,0){10}{\circle*{0.35}}
\put(48,70){\line(0,10){10}}
\put(55.6,74){\makebox(0,0)[r]{\small{$r$}}}
\put(58,88){\vector(0,1){2}}
\put(58,90){\vector(0,1){2}}
\put(58,74){\vector(0,1){3}}
\put(58,60){\line(0,10){30}}
\put(58,60){\vector(0,1){3}}
\put(58,58){\vector(0,1){3}}
\put(55.5,63){\makebox(0,0){\small{$n$}}}
\put(63.5,72){\makebox(0,0){\small{$k^{'}$}}}
\multiput(58,70)(1,0){10}{\circle*{0.35}}
\put(48,60){\line(0,1){10}}
\put(48,60){\vector(0,1){3}}
\put(45.5,63){\makebox(0,0){\small{$m$}}}
\put(48,58){\vector(0,1){3}}
\put(68,60){\line(0,1){20}}
\put(68,76){\vector(0,1){3}}
\put(68,78){\vector(0,1){3}}
\put(65.5,78){\makebox(0,0){\small{$m$}}}
\put(68,60){\vector(0,1){3}}
\put(68,58){\vector(0,1){3}}
\put(65.5,63){\makebox(0,0){\small{$p$}}}
\put(58,53){\makebox(0,0){$\text{A}_{2}$}}
\put(77,74){\makebox(0,0){\normalsize{\bf{+}}}}
\put(86,80){\line(0,1){10}}
\put(86,88){\vector(0,1){2}}
\put(86,90){\vector(0,1){2}}
\put(83.5,88){\makebox(0,0)[t]{\small{$n$}}}
\put(93.5,88){\makebox(0,0)[t]{\small{$p$}}}
\put(91,82){\makebox(0,0){\small{$k$}}}
\multiput(86,80)(1,0){10}{\circle*{0.35}}
\put(86,70){\line(0,10){10}}
\put(93.6,74){\makebox(0,0)[r]{\small{$r$}}}
\put(96,88){\vector(0,1){2}}
\put(96,90){\vector(0,1){2}}
\put(96,74){\vector(0,1){3}}
\put(96,60){\line(0,10){30}}
\put(96,60){\vector(0,1){3}}
\put(96,58){\vector(0,1){3}}
\put(93.5,63){\makebox(0,0){\small{$n$}}}
\put(101.5,72){\makebox(0,0){\small{$k^{'}$}}}
\multiput(96,70)(1,0){10}{\circle*{0.35}}
\put(86,60){\line(0,1){10}}
\put(86,60){\vector(0,1){3}}
\put(83.5,63){\makebox(0,0){\small{$m$}}}
\put(86,58){\vector(0,1){3}}
\put(106,60){\line(0,1){20}}
\put(106,80){\vector(0,1){3}}
\put(106,78){\vector(0,1){3}}
\put(103.5,78){\makebox(0,0){\small{$m$}}}
\put(106,60){\vector(0,1){3}}
\put(106,58){\vector(0,1){3}}
\put(103.5,63){\makebox(0,0){\small{$p$}}}
\put(96,53){\makebox(0,0){$\text{A}_{3}$}}
\put(115,74){\makebox(0,0){\normalsize{\bf{+}}}}
\put(124,80){\line(0,1){10}}
\put(124,88){\vector(0,1){2}}
\put(124,90){\vector(0,1){2}}
\put(121.5,88){\makebox(0,0)[t]{\small{$p$}}}
\put(131.5,88){\makebox(0,0)[t]{\small{$n$}}}
\put(129,82){\makebox(0,0){\small{$k$}}}
\multiput(124,80)(1,0){10}{\circle*{0.35}}
\put(124,70){\line(0,10){10}}
\put(131.6,74){\makebox(0,0)[r]{\small{$r$}}}
\put(134,88){\vector(0,1){2}}
\put(134,90){\vector(0,1){2}}
\put(134,74){\vector(0,1){3}}
\put(134,60){\line(0,10){30}}
\put(134,60){\vector(0,1){3}}
\put(134,58){\vector(0,1){3}}
\put(131.5,63){\makebox(0,0){\small{$p$}}}
\put(139.5,72){\makebox(0,0){\small{$k^{'}$}}}
\multiput(134,70)(1,0){10}{\circle*{0.35}}
\put(124,60){\line(0,1){10}}
\put(124,60){\vector(0,1){3}}
\put(121.5,63){\makebox(0,0){\small{$m$}}}
\put(124,58){\vector(0,1){3}}
\put(144,60){\line(0,1){20}}
\put(144,80){\vector(0,1){3}}
\put(144,78){\vector(0,1){3}}
\put(141.5,78){\makebox(0,0){\small{$m$}}}
\put(144,60){\vector(0,1){3}}
\put(144,58){\vector(0,1){3}}
\put(141.5,63){\makebox(0,0){\small{$n$}}}
\put(134,53){\makebox(0,0){$\text{A}_{4}$}}
\put(10,24){\makebox(0,0){\normalsize{\bf{=}}}}
\put(15,24){\makebox(0,0){\LARGE{\Bigg(}}}
\put(20,30){\line(0,1){10}}
\put(20,38){\vector(0,1){2}}
\put(20,40){\vector(0,1){2}}
\put(17.5,38){\makebox(0,0)[t]{\small{$n$}}}
\put(27.5,38){\makebox(0,0)[t]{\small{$p$}}}
\put(25,32){\makebox(0,0){\small{$k$}}}
\multiput(20,30)(1,0){10}{\circle*{0.35}}
\put(20,20){\line(0,10){10}}
\put(27.6,24){\makebox(0,0)[r]{\small{$r$}}}
\put(30,38){\vector(0,1){2}}
\put(30,40){\vector(0,1){2}}
\put(30,24){\vector(0,1){3}}
\put(30,10){\line(0,10){30}}
\put(30,10){\vector(0,1){3}}
\put(30,08){\vector(0,1){3}}
\put(27.5,13){\makebox(0,0){\small{$p$}}}
\put(35.5,22){\makebox(0,0){\small{$k^{'}$}}}
\multiput(30,20)(1,0){10}{\circle*{0.35}}
\put(20,10){\line(0,1){10}}
\put(20,10){\vector(0,1){3}}
\put(17.5,13){\makebox(0,0){\small{$m$}}}
\put(20,08){\vector(0,1){3}}
\put(40,10){\line(0,1){20}}
\put(40,30){\vector(0,1){3}}
\put(40,28){\vector(0,1){3}}
\put(37.5,28){\makebox(0,0){\small{$m$}}}
\put(40,10){\vector(0,1){3}}
\put(40,08){\vector(0,1){3}}
\put(37.5,13){\makebox(0,0){\small{$n$}}}
\put(30,03){\makebox(0,0){$\text{A}_{5}$}}
\put(49,24){\makebox(0,0){\normalsize{\bf{+}}}}
\put(58,30){\line(0,1){10}}
\put(58,38){\vector(0,1){2}}
\put(58,40){\vector(0,1){2}}
\put(55.5,38){\makebox(0,0)[t]{\small{$n$}}}
\put(65.5,38){\makebox(0,0)[t]{\small{$p$}}}
\put(63,32){\makebox(0,0){\small{$k$}}}
\multiput(58,30)(1,0){10}{\circle*{0.35}}
\put(58,20){\line(0,10){10}}
\put(65.6,24){\makebox(0,0)[r]{\small{$r$}}}
\put(68,38){\vector(0,1){2}}
\put(68,40){\vector(0,1){2}}
\put(68,24){\vector(0,1){3}}
\put(68,10){\line(0,10){30}}
\put(68,10){\vector(0,1){3}}
\put(68,08){\vector(0,1){3}}
\put(65.5,13){\makebox(0,0){\small{$n$}}}
\put(73.5,22){\makebox(0,0){\small{$k^{'}$}}}
\multiput(68,20)(1,0){10}{\circle*{0.35}}
\put(58,10){\line(0,1){10}}
\put(58,10){\vector(0,1){3}}
\put(55.5,13){\makebox(0,0){\small{$m$}}}
\put(58,08){\vector(0,1){3}}
\put(78,10){\line(0,1){20}}
\put(78,30){\vector(0,1){3}}
\put(78,28){\vector(0,1){3}}
\put(75.5,28){\makebox(0,0){\small{$m$}}}
\put(78,10){\vector(0,1){3}}
\put(78,08){\vector(0,1){3}}
\put(75.5,13){\makebox(0,0){\small{$p$}}}
\put(68,03){\makebox(0,0){$\text{A}_{6}$}}
\put(83,24){\makebox(0,0){\LARGE{\Bigg)}}}
\put(109,24){\makebox(0,0){\LARGE{\Bigg( 1 + P $
 \left( \hspace{-0.15cm}
\begin{array}{lcl}
      n&\hspace{-0.25cm}\rightleftharpoons&\hspace{-0.25cm}p \\
    \end{array}  \hspace{-0.15cm} \right)
$ \Bigg)}}}
\end{picture}
\caption{The VV Feynman diagrams of the second-order effective Hamiltonian, expressing the fourth type of valence-valence correlations 
$(n_{m} \ell_{m})\, j_{m}^{w_m} \, (n_{n} \ell_{n})\, j_{n}^{w_n} \, (n_{p} \ell_{p})\, j_{p}^{w_p} 
   \rightarrow (n_{m} \ell_{m})\, j_{m}^{w_m+1} \; (n_{n} \ell_{n})\, j_{n}^{w_n-1} \; (n_{p} \ell_{p})\, j_{p}^{w_p-1} \; (n_{r} \ell_{r})\, j_{r}$.
}
\label{VV_4_type}
\end{center}
\end{figure*}

The following type of correlation 
\begin{equation}
\label{eq:VVT-b} 
(n_{m} \ell_{m})\, j_{m}^{w_m} \, (n_{n} \ell_{n})\, j_{n}^{w_n} \, (n_{p} \ell_{p})\, j_{p}^{w_p} 
   \rightarrow (n_{m} \ell_{m})\, j_{m}^{w_m+1} \; (n_{n} \ell_{n})\, j_{n}^{w_n-1} \; (n_{p} \ell_{p})\, j_{p}^{w_p-1} \; (n_{r} \ell_{r})\, j_{r},
\end{equation}
is described by the same diagram VV$_3$ with four different sets of open lines with double-arrow indices. 
Four diagrams A$_1$, A$_2$, A$_3$, and A$_4$ from Fig.~\ref{VV_4_type}, represent all these different sets.
For example, Feynman diagram A$_1$ have the following values $m=p' \equiv n$, \, $m'=p \equiv m$, and $n=n' \equiv p$.
Therefore, to find the value of this type of correlation, it is necessary to analyze all four diagrams: A$_1$, A$_2$, A$_3$, and A$_4$. 
But it is easy to see that these diagrams can be converted into a search for two diagrams A$_5$ (A$_5 \equiv$ A$_1$), A$_6$ (A$_6 \equiv$ A$_3$) with a multiplier 
$\bigg( 1 + P \left( ... \right) \bigg)$, where the notation 
$P \left(  \hspace{-0.15cm}
\begin{array}{lcl}
      n&\hspace{-0.25cm}\rightleftharpoons&\hspace{-0.25cm}p \\
    \end{array}  \hspace{-0.15cm} \right)$
means that the diagrams need to change the index of $n$ to $p$ and $p$ to $n$.
This makes it considerably easier to carry out the desired calculations.
Note that A$_5$ describes the direct part of the correlation under consideration, while A$_6$ describes the exchange part of the interaction.
These A$_5$ and A$_6$ are three-particle Feynman diagram for which the program library \texttt{librang}~\cite{Gaigalas:2022} from {\sc Grasp}
does not support the calculation of spin-angular parts. It is, therefore, necessary to extend the capabilities of this program library to calculate the values of these diagrams. This problem will be discussed and solved in detail in the next section.

\section{The spin-angular part of the three-particle Feynman diagram contributing to valence-valence correlations}
\label{Sec:spin_angular}

The program library \texttt{librang}~\cite{Gaigalas:2022} from {\sc Grasp} evaluates spin-angular coefficients of any matrix element with any number of open subshells for any one- and/or two-particle operators. 
Therefore, in previous papers~\cite{Gaigetal:2024CV,Gaigetal:2024C,Gaigetal:2024CC,Gaigetal:2025VV}, when developing the combination of
second-order Rayleigh-Schr\"odinger perturbation theory
and relativistic configuration interaction method, there was no problem in using it to find correlations described by vacuum, one- and two-particle Feynman diagrams.
However, problems with using this library arise when dealing with three-particle operators.

The program library \texttt{librang}~\cite{Gaigalas:2022} is based on the spin-angular approach~\cite{Gaigalas_1996,Gaigalas_1997}. The specifics of this approach (factorization of standard values, interaction strengths, and recoupling matrices) make it easy enough to extend it to allow the program library to find the spin-angular part of the VV$_3$ Feynman diagram that is needed to find the correlations that are being studied in this paper. This can be done by using ideas published in the paper~\cite{Gaigalas:89}, where the second quantization operators are grouped according to their action on the subshell, i.e., so that the second quantization operators acting on the subshell $m$ first, then on the subshell $p$ (if there are any), and finally on the subshell $n$. Below, we show how this has been done for the different types of correlations, separately.

\subsection{The spin-angular part of the third type of valence-valence correlations}
\label{sec:PT_SA_Third}

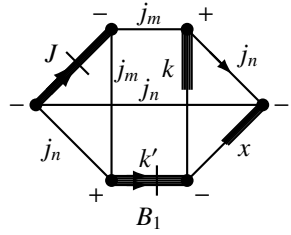
\begin{figure}
\begin{center}
\setlength{\unitlength}{1mm}
\begin{picture}(47,47)
\thicklines
\put(12,27){\makebox(0,0)[t]{$J$}}
\put(38,27){\makebox(0,0)[t]{$j_{n}$}}
\put(22,25){\makebox(0,0)[t]{$j_{m}$}}
\put(27.5,25){\makebox(0,0)[t]{$k$}}
\put(20,29){\circle*{1.7}}
\put(18.5,31){\makebox(0,0){$-$}}
\put(30,29){\circle*{1.7}}
\put(32.5,31){\makebox(0,0){$+$}}
\put(20,29){\line(10,0){10}}
\put(25,31){\makebox(0,0){$j_{m}$}}
\put(10,19){\circle*{1.7}}
\put(40,19){\circle*{1.7}}
\put(10,19){\line(30,0){30}}
\put(33,26){\vector(1,-1){3}}
\put(25,21){\makebox(0,0){$j_{n}$}}
\put(20,09){\circle*{1.7}}
\put(18,07){\makebox(0,0){$+$}}
\put(30,09){\circle*{1.7}}
\put(32,07){\makebox(0,0){$-$}}
\put(20,09.6){\line(10,0){10}}
\put(20,09.3){\line(10,0){10}}
\put(20,09){\line(10,0){10}}
\put(20,08.7){\line(10,0){10}}
\put(20,08.4){\line(10,0){10}}
\put(26,07){\line(0,4){4}}
\put(22,09.6){\vector(1,0){3}}
\put(22,08.4){\vector(1,0){3}}
\put(25,12){\makebox(0,0){$k'$}}
\put(9.4,19){\line(1,1){10}}
\put(9.7,19){\line(1,1){10}}
\put(10,19){\line(1,1){10}}
\put(10.3,19){\line(1,1){10}}
\put(10.6,19){\line(1,1){10}}
\put(17,23){\line(-2,2){3}}
\put(11.4,21.0){\vector(1,1){3}}
\put(12,20.4){\vector(1,1){3}}
\put(10,19){\line(1,-1){10}}
\put(39.4,19){\line(-1,-1){4.8}}
\put(39.7,19){\line(-1,-1){4.9}}
\put(40,19){\line(-1,1){10}}
\put(40,19){\line(-1,-1){10}}
\put(40.3,19){\line(-1,-1){5.2}}
\put(40.7,19){\line(-1,-1){5.3}}
\put(20,09){\line(0,20){20}}
\put(12,13){\makebox(0,0){$j_{n}$}}
\put(07.5,19){\makebox(0,0){$-$}}
\put(29.4,29){\line(0,-8){8}}
\put(29.7,29){\line(0,-8){8}}
\put(30,09){\line(0,20){20}}
\put(30.3,29){\line(0,-8){8}}
\put(30.6,29){\line(0,-8){8}}
\put(37.5,13){\makebox(0,0){$x$}}
\put(42.5,19){\makebox(0,0){$-$}}
\put(25,04){\makebox(0,0){$ B_{1}$}}
\end{picture}
\caption{Diagram showing the recoupling coefficient $B_1$.}
\label{B1}
\end{center}
\end{figure}

\begin{figure}
\begin{center}
\setlength{\unitlength}{1mm}
\begin{picture}(47,47)
\thicklines
\put(12,27){\makebox(0,0)[t]{$k$}}
\put(38,27){\makebox(0,0)[t]{$j_{m}$}}
\put(20.5,24.5){\makebox(0,0)[t]{$x$}}
\put(29,25){\makebox(0,0)[t]{$j_n$}}
\put(20,29){\circle*{1.7}}
\put(18.5,31){\makebox(0,0){$+$}}
\put(30,29){\circle*{1.7}}
\put(32.5,31){\makebox(0,0){$-$}}
\put(20,29.6){\line(10,0){10}}
\put(20,29.3){\line(10,0){10}}
\put(20,29){\line(10,0){10}}
\put(20,28.7){\line(10,0){10}}
\put(20,28.4){\line(10,0){10}}
\put(24,27){\line(0,4){4}}
\put(28,29.6){\vector(-1,0){3}}
\put(28,28.4){\vector(-1,0){3}}
\put(25,32.5){\makebox(0,0){$k'$}}
\put(10,19){\circle*{1.7}}
\put(40,19){\circle*{1.7}}
\put(10.5,19){\line(29,0){30}}
\put(17,21){\makebox(0,0){$j_{m}$}}
\put(20,09){\circle*{1.7}}
\put(18,07){\makebox(0,0){$-$}}
\put(30,09){\circle*{1.7}}
\put(32,07){\makebox(0,0){$+$}}
\put(20,09.6){\line(5,0){5}}
\put(20,09.3){\line(5,0){5}}
\put(20,09){\line(10,0){10}}
\put(20,08.7){\line(5,0){5}}
\put(20,08.4){\line(5,0){5}}
\put(25,13){\makebox(0,0){$J_{2}$}}
\put(9.4,19){\line(1,1){5.3}}
\put(9.7,19){\line(1,1){5.2}}
\put(10,19){\line(1,1){10}}
\put(10.3,19){\line(1,1){4.9}}
\put(10.6,19){\line(1,1){4.8}}
\put(10,19){\line(1,-1){10}}
\put(40,19){\line(-1,1){10}}
\put(39.4,19){\line(-1,-1){10}}
\put(39.7,19){\line(-1,-1){10}}
\put(40,19){\line(-1,-1){10}}
\put(40.3,19){\line(-1,-1){10}}
\put(40.7,19){\line(-1,-1){10}}
\put(37,13){\line(-2,2){3}}
\put(31.4,11.0){\vector(1,1){3}}
\put(32,10.4){\vector(1,1){3}}
\put(20,09){\line(1,2){10}}
\put(12,13){\makebox(0,0){$j_{n}$}}
\put(07.5,19){\makebox(0,0){$+$}}
\put(30,09){\line(-1,2){10}}
\put(38.5,13){\makebox(0,0){$J_{1}$}}
\put(42.5,19){\makebox(0,0){$-$}}
\put(25,04){\makebox(0,0){$ B_{2}$}}
\end{picture}
\caption{Diagram showing the recoupling coefficient $B_2$ of the tensorial product from (\ref{eq:Tensor11}) to (\ref{eq:Tensor12}).}
\label{B2}
\end{center}
\end{figure}
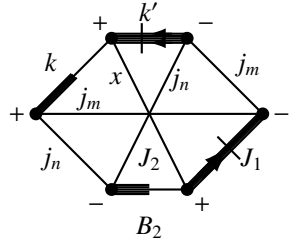

In this case, the diagram VV$_3$ has the following tensorial structure
\begin{equation}
\label{eq:Tensor11}
\biggl[ \Bigl[ \bigl[ a^{(j_n)}_1 \times \tilde{a}^{(j_m)}_2 \bigr]^{(k)}   \times \bigl[ a^{(j_n)}_3 \times \tilde{a}^{(j_n)}_4 \bigr]^{(x)} \Bigr]^{(k')}  \times \bigl[ a^{(j_m)}_5 \times \tilde{a}^{(j_n)}_6 \bigr]^{(k')} \biggr]^{(0)},
\end{equation}
where the subscript next to the operator indicates the sequence number of the second quantization operator in the tensorial product.
As seen in the tensorial structure (\ref{eq:Tensor11}), the two pairs of second quantization operators are mixed, i.e., they consist of second quantization operators acting on different subshells. This is a pair consisting of operators with sequence numbers 1 and 2 and a pair with sequence numbers 5 and 6. This situation greatly complicates the study of the spin-angular part of this operator. But if we replace the tensorial structure (\ref{eq:Tensor11}) with the following
\begin{equation}
\label{eq:Tensor12}
\biggl[ \bigl[ \tilde{a}^{(j_m)}_2 \times a^{(j_m)}_5 \bigr]^{(J_1)} \times \Bigl[ \bigl[ a^{(j_n)}_3 \times \tilde{a}^{(j_n)}_4 \bigr]^{(x)} \times \bigl[ a^{(j_n)}_1 \times \tilde{a}^{(j_n)}_6 \bigr]^{(J_2)}  \Bigr]^{(J_1)} \biggr]^{(0)}
\end{equation}
then the situation would change and one could easily extend the spin-angular approach~\cite{Gaigalas_1996,Gaigalas_1997}
to the study of three-particle operator for this type of correlations~\cite{Gaigalas:89}.
This can be done by using the commutation rule of second quantization operators.
This rule allows us to transform the tensorial product (\ref{eq:Tensor11}) into (\ref{eq:Tensor12}) using commutations, resulting in a suitable form for calculation in which we have a linear combination of two sets of tensorial operators. The first tensorial product 
$\bigl[ \tilde{a}^{(j_m)}_2 \times a^{(j_m)}_5 \bigr]^{(J_1)}$
is the one to which we want to bring the algebraic expression, i.e. (\ref{eq:Tensor12}), and the second one 
$\Bigl[ \bigl[ a^{(j_n)}_3 \times \tilde{a}^{(j_n)}_4 \bigr]^{(x)} \times \bigl[ a^{(j_n)}_1 \times \tilde{a}^{(j_n)}_6 \bigr]^{(J_2)}  \Bigr]^{(J_1)}$
consists of only two pairs of operators for the second quantization acting to the same subshell. These two sets of tensorial operators are obtained by applying the operator commutation rule to operators $a^{(j_n)}_1$ and $\tilde{a}^{(j_n)}_4$.
This gives the following expression
\begin{equation}
\label{eq:Tensor13}
\begin{aligned}
\biggl[ \Bigl[ \bigl[ a^{(j_n)} \times \tilde{a}^{(j_m)} \bigr]^{(k)}   \times \bigl[ a^{(j_n)} \times \tilde{a}^{(j_n)} \bigr]^{(x)} \Bigr]^{(k')}  \times \bigl[ a^{(j_m)} \times \tilde{a}^{(j_n)} \bigr]^{(k')} \biggr]^{(0)}
\\
= \;
\sum_{J} B_1 \; \biggl[ \bigl[ \tilde{a}^{(j_m)} \times a^{(j_m)} \bigr]^{(J)} \times \bigl[ a^{(j_n)} \times \tilde{a}^{(j_n)} \bigr]^{(J)} \biggr]^{(0)}
\\
+ \;
\sum_{J_1,J_2} B_2 \; \biggl[ \bigl[ \tilde{a}^{(j_m)} \times a^{(j_m)} \bigr]^{(J_1)} \times \Bigl[ \bigl[ a^{(j_n)} \times \tilde{a}^{(j_n)} \bigr]^{(x)}  \times \bigl[ a^{(j_n)} \times \tilde{a}^{(j_n)} \bigr]^{(J_2)} \Bigr]^{(J_1)} \biggr]^{(0)}.
\end{aligned}
\end{equation}
The coefficients $B_1$ and $B_2$ in the expression (\ref{eq:Tensor13}) are the easiest to represent and their algebraic expressions are the easiest to obtain by using the generalized graphical method of the angular momentum theory~\cite{Gaigatal:85}. In this graphical representation, the $B_1$ multiplier is shown graphically in Fig.~\ref{B1} and the $B_2$ multiplier in Fig.~\ref{B2}.

Using the Diagrams Separation on Three Lines Theorem of Jucys, Levinson, and Vanagas (see \cite{LindgrenBook:82} Section 4.1.3), it is possible to split diagram $B_1$ into two diagrams, each representing a 6$j$-symbol~\cite{Gaigatal:85,JucBan:77a}. This rule is general and is used in all versions of the graphical methods for the angular momentum, whether the graphical method is applied to the Wigner coefficients~\cite{LindgrenBook:82,Yutetal:62a,BriSat:68,BlaCas:72,Varshalovich:2021} or to the Clebsch-Gordan coefficients~\cite{Gaigatal:85,JucBan:77a,BlaCas:72}. In this case, the diagram $B_1$ cuts through the horizontal lines $j_{m}$, $j_{n}$ , and $k'$. Meanwhile, the diagram $B_2$ uses a graphical technique~\cite{Gaigatal:85,JucBan:77a} to represent the 9$j$-symbol with additional multipliers. The Eq.~(\ref{eq:Tensor13}) can be rewritten as
\begin{equation}
\label{eq:Tensor14}
\begin{aligned}
\biggl[ \Bigl[ \bigl[ a^{(j_n)} \times \tilde{a}^{(j_m)} \bigr]^{(k)}   \times \bigl[ a^{(j_n)} \times \tilde{a}^{(j_n)} \bigr]^{(x)} \Bigr]^{(k')}  \times \bigl[ a^{(j_m)} \times \tilde{a}^{(j_n)} \bigr]^{(k')} \biggr]^{(0)}
\\
= \; \left( -1 \right)^{j_m+j_n+k+x} \sqrt{\left[ k, k', x \right]}
\left\{
    \begin{array}{ccc}
      j_m & j_n & k \\
      x   & k'  & j_{n}
    \end{array}
\right\}
\sum_{J} \left( -1 \right)^{J} \sqrt{\left[ J \right]}
\left\{
    \begin{array}{ccc}
      j_m & j_m & J \\
      j_n & j_n & k'
    \end{array}
\right\}
\\
\times \; \biggl[ \bigl[ \tilde{a}^{(j_m)} \times a^{(j_m)} \bigr]^{(J)} \times \bigl[ a^{(j_n)} \times \tilde{a}^{(j_n)} \bigr]^{(J)} \biggr]^{(0)}
\\
+ \;
\left( -1 \right)^{j_m+j_n+k+x} \sqrt{\left[ k, k' \right]} \sum_{J_1,J_2} \sqrt{\left[ J_1, J_2 \right]}
\left\{
    \begin{array}{ccc}
      j_m & j_n & k \\
			J_1 & J_2 & x \\
      j_m & j_n & k'
    \end{array}
\right\}
\\
\times \; \biggl[ \bigl[ \tilde{a}^{(j_m)} \times a^{(j_m)} \bigr]^{(J_1)} \times \Bigl[ \bigl[ a^{(j_n)} \times \tilde{a}^{(j_n)} \bigr]^{(x)} \times \bigl[ a^{(j_n)} \times \tilde{a}^{(j_n)} \bigr]^{(J_2)}  \Bigr]^{(J_1)} \biggr]^{(0)} .
\end{aligned}
\end{equation}

The first term in the expression (\ref{eq:Tensor14}) corresponds to the spin-angular part of the two-particle operators. The program library \texttt{librang}~\cite{Gaigalas:2022} is therefore sufficient for its calculation. The second term, with its three pairs of second quantization operators, corresponds to the three-particle operator (\ref{eq:Tensor12}). In this case, adding new features to the library is necessary, which we will now discuss.

A peculiarity of the methodology~\cite{Gaigalas_1997} is that in the expression~\cite[(11)]{Gaigalas_1997} the recoupling matrix $R\left( \lambda _i,\lambda _j,\lambda
_i^{\prime },\lambda _j^{\prime },\Lambda ^{bra},\Lambda ^{ket},\Gamma
\right) $, the submatrix element $T\left( n_i\lambda _i,n_j\lambda _j,n_i^{\prime }\lambda _i^{\prime
},n_j^{\prime }\lambda _j^{\prime },\Lambda ^{bra},\Lambda ^{ket},\Xi
,\Gamma \right) $, the phase factor $\Delta $, and $\Theta ^{\prime }\left( n_i\lambda _i,n_j\lambda _j,n_i^{\prime
}\lambda _i^{\prime },n_j^{\prime }\lambda _j^{\prime },\Xi \right)$, which is proportional to the radial part, are easily separated from each other and can be treated differently.
This fact makes the methodology flexible and allows it to be easily extended to include new class/type operators, such as the three-particle operator (\ref{eq:Tensor12}) we are considering. In particular, how the expression ~\cite[(11)]{Gaigalas_1997} is implemented in the library is described in the paper~\cite{Gaigalas:2022}. Here, we will now discuss only those aspects of the tensorial product (\ref{eq:Tensor12}) of Feynman diagram $A_5$ which was not covered in the paper~\cite{Gaigalas:2022} and show which subroutines from the program library \texttt{librang} should be used and which expressions should be added to it.

Since we can schematically rewrite the tensorial product as the tensorial product of two operators $A^{(J_{1})}(n_{m}j_{m})$ and $B^{(J_{1})}(n_{n}j_{n})$ acting on different subshells
\begin{equation}
\label{eq:tena}
\left[\; A^{(J_{1})}(n_{m}j_{m}) \times B^{(J_{1})}(n_{n}j_{n}) \; \right]^{(0)},
\end{equation}
the algebraic expression of the recoupling matrix can be used from the paper~\cite[(19)]{Gaigalas_1997} and can be computed by the subroutine \texttt{RECO2}~\cite[ Section 3.2.4]{Gaigalas:2022}. Therefore, the library's \texttt{librang} existing capabilities are entirely sufficient for calculating $R\left( \lambda _i,\lambda _j,\lambda
_i^{\prime },\lambda _j^{\prime },\Lambda ^{bra},\Lambda ^{ket},\Gamma
\right) $.

As far as the submatrix element $T\left( n_i\lambda _i,n_j\lambda _j,n_i^{\prime }\lambda _i^{\prime
},n_j^{\prime }\lambda _j^{\prime },\Lambda ^{bra},\Lambda ^{ket},\Xi
,\Gamma \right) $ is concerned, the situation is different. In this case, the tensorial product can be split into two parts, i.e., 
\begin{equation}
\label{eq:tenb}
A^{(J_{1})}(n_{m}j_{m}) \; \equiv \bigl[ \tilde{a}^{(j_m)} \times a^{(j_m)} \bigr]^{(J_1)}
\end{equation}
and 
\begin{equation}
\label{eq:tenc}
B^{(J_{1})}(n_{n}j_{n}) \;  \equiv \Bigl[ \bigl[ a^{(j_n)} \times \tilde{a}^{(j_n)} \bigr]^{(x)}  \times \bigl[ a^{(j_n)} \times \tilde{a}^{(j_n)} \bigr]^{(J_2)} \Bigr]^{(J_1)}.
\end{equation}
These two members should be considered separately (because they act on different subshells, and the binding of the ranks of the tensorial structure is already included in the recoupling matrix). Subroutine  \texttt{WJ1} \cite[Section 3.3.3]{Gaigalas:2022} finds the matrix element of operator (\ref{eq:tenb}), while the library \texttt{librang} has no suitable subroutine for computing the matrix element of the operator (\ref{eq:tenc}). To find the latter requires the use of an expression such as 
\begin{eqnarray}
\label{eq:tg}
\hspace{-2.0cm}
   \redmem{(n_n\ell_n)\, j_n^w\, \alpha J}
	        {\, \Bigl[ \bigl[ a^{(j_n)} \times \tilde{a}^{(j_n)} \bigr]^{(x)}  \times \bigl[ a^{(j_n)} \times \tilde{a}^{(j_n)} \bigr]^{(J_2)} \Bigr]^{(J_1)} \,}
					{(n_n\ell_n)\, j_n^{w} \, \alpha ^{\prime }J^{\prime }}
   \nonumber  \\[1ex]
\hspace{-1.5cm}
   =\left( -1\right) ^{J + J^{\prime } + J_1} \ \sqrt{\left[ J_1 \right]} \ 
   \displaystyle {\sum_{\alpha ^{\prime \prime }J^{\prime \prime }}}
   \ \left\{
   \begin{array}{ccc}
      x         & J_2 & J_1 \\
      J^{\prime } & J   & J^{\prime \prime }
   \end{array}
   \right\}
   \nonumber  \\[1ex]
\hspace{-1cm}
   \times
   \redmem{(n_n\ell_n)\, j_n^w \,\alpha J}{\, \bigl[ a^{(j_n)} \times \tilde{a}^{(j_n)} \bigr]^{(x)} \,}
	        {(n_n\ell_n)\, j_n^{w}\, \alpha ^{\prime \prime }J^{\prime \prime }} \
   \nonumber  \\[1ex]
\hspace{-1cm}
   \times
   \redmem{(n_n\ell_n)\, j_n^{w}\, \alpha ^{\prime \prime }J^{\prime \prime }}
	        {\, \bigl[ a^{(j_n)} \times \tilde{a}^{(j_n)} \bigr]^{(J_2)} \,}{(n_n\ell_n)\, j_n^{w}\, \alpha ^{\prime }J^{\prime }}.
\end{eqnarray}
Therefore, the library \texttt{librang} must be extended by programming the expression (\ref{eq:tg}).

The phase factor $\Delta $, according to \cite{Gaigalas_1997}, is zero.

The $\Theta ^{\prime }\left( n_i\lambda _i,n_j\lambda _j,n_i^{\prime
}\lambda _i^{\prime },n_j^{\prime }\lambda _j^{\prime },\Xi \right)$, which is proportional to the radial part, is found in a regular way as it was found in papers~\cite{Gaigetal:2024CV,Gaigetal:2024C,Gaigetal:2024CC,Gaigetal:2025VV}.

Note that the above describes how to calculate the reduced matrix elements of Feynman diagram VV$_3$ in the general case for the third type of valence-valence correlations. But it is possible to extract the individual parts which are more straightforward to calculate, i.e. the coefficient $\Delta \mathcal{E}_0$ (does not depend on the term), $\Delta \mathcal{F}^{k}(n,n)$, and $\Delta \mathcal{F}^{k}(m,n)$ (regular spin-angular library \texttt{librang} can be used). The extraction of these expressions is the same as in papers~\cite{Gaigetal:2024CV,Gaigetal:2024C,Gaigetal:2024CC,Gaigetal:2025VV}. Therefore, only the reduced matrix element of part of Feynman diagrams $A_5$ with tensorial product (\ref{eq:Tensor12}) is calculated from the general expression when the rank $k > 0$. The part which is calculated from the general expression will be denoted by $\Delta \widetilde{\mathcal{R}}^{(k, k', x)} \left( m n n  \right)$ in the future.

\subsection{The spin-angular part of the fourth type of valence-valence correlations}
\label{sec:PT_SA_Fourth}

Two different Feynman diagrams $A_5$ and $A_6$ from Fig.~\ref{VV_4_type} describe this type of correlation. We will consider each separately, as their spin-angular part differs significantly.

\subsubsection{The direct part of the fourth type of valence-valence correlations}
\label{sec:PT_SA_direct_Fourth}

Now let us discuss the diagram $A_5$, with tensorial structure
\begin{equation}
\label{eq:Tensor211}
\biggl[ \Bigl[ \bigl[ a^{(j_n)}_1 \times \tilde{a}^{(j_m)}_2 \bigr]^{(k)}   \times \bigl[ a^{(j_p)}_3 \times \tilde{a}^{(j_p)}_4 \bigr]^{(x)} \Bigr]^{(k')}  \times \bigl[ a^{(j_m)}_5 \times \tilde{a}^{(j_n)}_6 \bigr]^{(k')} \biggr]^{(0)}.
\end{equation}
As in (\ref{eq:Tensor11}), (\ref{eq:Tensor211}) is not appropriate to calculate the spin-angular part because there are two pairs of secondary quantization operators acting on different layers, i.e., $\bigl[ a^{(j_n)}_1 \times \tilde{a}^{(j_m)}_2 \bigr]^{(k)}$ and $\bigl[ a^{(j_m)}_5 \times \tilde{a}^{(j_n)}_6 \bigr]^{(k)}$. Therefore, this tensorial product has to be transformed into the following, using the commutation rules of secondary quantization
\begin{equation}
\label{eq:Tensor212}
\biggl[ \Bigl[ \bigl[ \tilde{a}^{(j_m)}_2 \times a^{(j_m)}_5 \bigr]^{(J_1)} \times  \bigl[ a^{(j_p)}_3 \times \tilde{a}^{(j_p)}_4 \bigr]^{(x)} \Bigr]^{(J_2)} \times \bigl[ a^{(j_n)}_1 \times \tilde{a}^{(j_n)}_6 \bigr]^{(J_2)}  \biggr]^{(0)}.
\end{equation}
\begin{figure}
\begin{center}
\setlength{\unitlength}{1mm}
\begin{picture}(47,47)
\thicklines
\put(12,27){\makebox(0,0)[t]{$k$}}
\put(38,27){\makebox(0,0)[t]{$j_{m}$}}
\put(20.5,24.5){\makebox(0,0)[t]{$x$}}
\put(29,25){\makebox(0,0)[t]{$j_n$}}
\put(20,29){\circle*{1.7}}
\put(18.5,31){\makebox(0,0){$+$}}
\put(30,29){\circle*{1.7}}
\put(32.5,31){\makebox(0,0){$-$}}
\put(20,29.6){\line(10,0){10}}
\put(20,29.3){\line(10,0){10}}
\put(20,29){\line(10,0){10}}
\put(20,28.7){\line(10,0){10}}
\put(20,28.4){\line(10,0){10}}
\put(24,27){\line(0,4){4}}
\put(28,29.6){\vector(-1,0){3}}
\put(28,28.4){\vector(-1,0){3}}
\put(25,32.5){\makebox(0,0){$k'$}}
\put(10,19){\circle*{1.7}}
\put(40,19){\circle*{1.7}}
\put(10.5,19){\line(29,0){30}}
\put(17,21){\makebox(0,0){$j_{m}$}}
\put(20,09){\circle*{1.7}}
\put(18,07){\makebox(0,0){$-$}}
\put(30,09){\circle*{1.7}}
\put(32,07){\makebox(0,0){$+$}}
\put(20,09.6){\line(10,0){10}}
\put(20,09.3){\line(10,0){10}}
\put(20,09){\line(10,0){10}}
\put(20,08.7){\line(10,0){10}}
\put(20,08.4){\line(10,0){10}}
\put(26,07){\line(0,4){4}}
\put(22,09.6){\vector(1,0){3}}
\put(22,08.4){\vector(1,0){3}}
\put(25,13){\makebox(0,0){$J_{2}$}}
\put(9.4,19){\line(1,1){5.3}}
\put(9.7,19){\line(1,1){5.2}}
\put(10,19){\line(1,1){10}}
\put(10.3,19){\line(1,1){4.9}}
\put(10.6,19){\line(1,1){4.8}}
\put(10,19){\line(1,-1){10}}
\put(40,19){\line(-1,1){10}}
\put(39.4,19){\line(-1,-1){4.8}}
\put(39.7,19){\line(-1,-1){4.9}}
\put(40,19){\line(-1,-1){10}}
\put(40.3,19){\line(-1,-1){5.2}}
\put(40.7,19){\line(-1,-1){5.3}}
\put(20,09){\line(1,2){10}}
\put(12,13){\makebox(0,0){$j_{n}$}}
\put(07.5,19){\makebox(0,0){$+$}}
\put(30,09){\line(-1,2){10}}
\put(38,13){\makebox(0,0){$J_{1}$}}
\put(42.5,19){\makebox(0,0){$-$}}
\put(25,04){\makebox(0,0){$ B_{3}$}}
\end{picture}
\caption{Diagram showing the recoupling coefficient $B_3$ of the tensorial product from (\ref{eq:Tensor211}) to (\ref{eq:Tensor212}).}
\label{B3}
\end{center}
\end{figure}
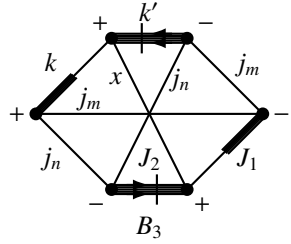
In this case, the transformation of the tensorial product from (\ref{eq:Tensor211}) to (\ref{eq:Tensor212}) results in only one member in (\ref{eq:Tensor212}), where the transformation matrix of the tensorial structure is shown graphically in Fig.~\ref{B3}. $B_3$ is proportional to the coefficient 9$j$-symbol~\cite{Gaigatal:85,JucBan:77a}. The final expression of this rearrangement of the tensorial structure (\ref{eq:Tensor211}) is the following:
\begin{equation}
\label{eq:Tensor213}
\begin{aligned}
\biggl[ \Bigl[ \bigl[ a^{(j_n)} \times \tilde{a}^{(j_m)} \bigr]^{(k)}   \times \bigl[ a^{(j_p)} \times \tilde{a}^{(j_p)} \bigr]^{(x)} \Bigr]^{(k')}  \times \bigl[ a^{(j_m)} \times \tilde{a}^{(j_n)} \bigr]^{(k')} \biggr]^{(0)}
\\
= \;
\sum_{J_1,J_2} B_3 \; \biggl[ \Bigl[ \bigl[ \tilde{a}^{(j_m)} \times a^{(j_m)} \bigr]^{(J_1)} \times \bigl[ a^{(j_p)} \times \tilde{a}^{(j_p)} \bigr]^{(x)} \Bigr]^{(J_2)}  \times \bigl[ a^{(j_n)} \times \tilde{a}^{(j_n)} \bigr]^{(J_2)} \biggr]^{(0)}
\end{aligned}
\end{equation}
and
\begin{equation}
\label{eq:Tensor214}
\begin{aligned}
\biggl[ \Bigl[ \bigl[ a^{(j_n)} \times \tilde{a}^{(j_m)} \bigr]^{(k)}   \times \bigl[ a^{(j_p)} \times \tilde{a}^{(j_p)} \bigr]^{(x)} \Bigr]^{(k')}  \times \bigl[ a^{(j_m)} \times \tilde{a}^{(j_n)} \bigr]^{(k')} \biggr]^{(0)}
\\
= \;
\left( -1 \right)^{j_m+j_n+k+x} \sqrt{\left[ k, k' \right]} \sum_{J_1,J_2} \sqrt{\left[ J_1, J_2 \right]}
\left\{
    \begin{array}{ccc}
      j_m & j_n & k \\
			J_1 & J_2 & x \\
      j_m & j_n & k'
    \end{array}
\right\}
\\
\times \; \biggl[ \Bigl[ \bigl[ \tilde{a}^{(j_m)} \times a^{(j_m)} \bigr]^{(J_1)} \times \bigl[ a^{(j_p)} \times \tilde{a}^{(j_p)} \bigr]^{(x)} \Bigr]^{(J_2)} \times \bigl[ a^{(j_n)} \times \tilde{a}^{(j_n)} \bigr]^{(J_2)}  \biggr]^{(0)}.
\end{aligned}
\end{equation}
The program library \texttt{librang}~\cite{Gaigalas:2022} available in the {\sc Grasp} package is sufficient for the calculation of the reduced matrix element of Feynman diagrams $A_5$, but the method of calculation of this type of operator is not described in the paper \cite{Gaigalas:2022}. We will now discuss it.

Since we can schematically rewrite the tensorial product as the tensorial product of three operators $A^{(J_1)}(n_{m}j_{m})$, $B^{(x)}(n_{p}j_{p})$,  and $C^{(J_2)}(n_{m}j_{m})$ acting on different subshells
\begin{equation}
\label{eq:tend}
\left[ \left[ \; A^{(J_{1})}(n_{m}j_{m}) \times B^{(x)}(n_{p}j_{p}) \; \right]^{(k)} \times
C^{(J_{2})}(n_{m}j_{m}) \; \right]^{(0)},
\end{equation}
the algebraic expression of the recoupling matrix in this case is given in the paper~\cite[(24)]{Gaigalas_1997} and is computed by the subroutine \texttt{REC3}~\cite[ Section 3.2.5]{Gaigalas:2022}. Therefore, the library's \texttt{librang} existing capabilities are fully sufficient for calculating $R\left( \lambda _i,\lambda _j,\lambda
_i^{\prime },\lambda _j^{\prime },\Lambda ^{bra},\Lambda ^{ket},\Gamma
\right)$.

As far as the submatrix element $T\left( n_i\lambda _i,n_j\lambda _j,n_i^{\prime }\lambda _i^{\prime
},n_j^{\prime }\lambda _j^{\prime },\Lambda ^{bra},\Lambda ^{ket},\Xi
,\Gamma \right) $ is concerned, the situation is different. In this case, the tensorial product can be split into three parts $A^{(J_{1})}(n_{m}j_{m})$, $B^{(x)}(n_{p}j_{p})$, and $C^{(J_{2})}(n_{m}j_{m})$, where each of these can be expressed as
\begin{equation} 
\label{eq:tene}
A^{(J_{1})}(n_{m}j_{m}) = \bigl[ \tilde{a}^{(j_m)} \times a^{(j_m)} \bigr]^{(J_{1})}
\end{equation}
and
\begin{equation} 
\label{eq:tenf}
B^{(k)}(nj) = C^{(k)}(nj) = \bigl[ a^{(j)} \times \tilde{a}^{(j)} \bigr]^{(k)}.
\end{equation}
These parts can be considered separately (because they act on different subshells, and the binding of the ranks of the tensorial structure is already included in the recoupling matrix). Subroutine  \texttt{WJ1} \cite[Section 3.3.3]{Gaigalas:2022} finds the matrix element of operator (\ref{eq:tend}). 

The phase factor $\Delta $ is zero in this case.

The $\Theta ^{\prime }\left( n_i\lambda _i,n_j\lambda _j,n_i^{\prime
}\lambda _i^{\prime },n_j^{\prime }\lambda _j^{\prime },\Xi \right)$, which is proportional to the radial part, is found in a regular way as it was found in the papers~\cite{Gaigetal:2024CV,Gaigetal:2024C,Gaigetal:2024CC,Gaigetal:2025VV}.

\subsubsection{The exchange part of the fourth type of valence-valence correlations}
\label{sec:PT_SA_exchange_Fourth}

Now let us discuss the diagram $A_6$, with tensorial structure
\begin{equation}
\label{eq:Tensor311}
\biggl[ \Bigl[ \bigl[ a^{(j_n)}_1 \times \tilde{a}^{(j_m)}_2 \bigr]^{(k)}   \times \bigl[ a^{(j_p)}_3 \times \tilde{a}^{(j_n)}_4 \bigr]^{(x)} \Bigr]^{(k')}  \times \bigl[ a^{(j_m)}_5 \times \tilde{a}^{(j_p)}_6 \bigr]^{(k')} \biggr]^{(0)}.
\end{equation}
This tensorial product is not appropriate to calculate the spin-angular part for the same reason as for (\ref{eq:Tensor11}) and (\ref{eq:Tensor211}). Therefore, this tensorial product has to be transformed into the following, using the commutation rules of secondary quantization
\begin{equation}
\label{eq:Tensor312}
\biggl[ \Bigl[ \bigl[ \tilde{a}^{(j_m)}_2 \times a^{(j_m)}_5 \bigr]^{(J_1)} \times  \bigl[ a^{(j_p)}_3 \times \tilde{a}^{(j_p)}_6 \bigr]^{(y)} \Bigr]^{(J_2)} \times \bigl[ a^{(j_n)}_1 \times \tilde{a}^{(j_n)}_4 \bigr]^{(J_2)}  \biggr]^{(0)}.
\end{equation}
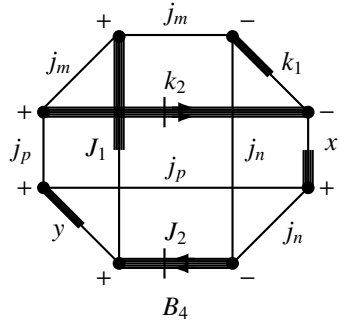
\begin{figure}
\begin{center}
\setlength{\unitlength}{1mm}
\begin{picture}(45,45)
\thicklines
\put(12,37){\makebox(0,0)[t]{$j_{m}$}}
\put(43,37){\makebox(0,0)[t]{$k_{1}$}}
\put(17,25.5){\makebox(0,0)[t]{$J_{1}$}}
\put(38,25.5){\makebox(0,0)[t]{$j_{n}$}}
\put(20,39){\circle*{1.7}}
\put(18,41){\makebox(0,0){$\bf{+}$}}
\put(35,39){\circle*{1.7}}
\put(37,41){\makebox(0,0){$\bf{-}$}}
\put(20,39.2){\line(15,0){15}}
\put(27.5,42){\makebox(0,0){$j_m$}}
\put(10,29){\circle*{1.7}}
\put(45,29){\circle*{1.7}}
\put(10,29.6){\line(35,0){35}}
\put(10,29.3){\line(35,0){35}}
\put(10,29){\line(35,0){35}}
\put(10,28.7){\line(35,0){35}}
\put(10,28.4){\line(35,0){35}}
\put(26,27){\line(0,4){4}}
\put(27,29.6){\vector(1,0){3}}
\put(27,28.4){\vector(1,0){3}}
\put(27.5,33){\makebox(0,0){$k_{2}$}}
\put(10,19){\circle*{1.7}}
\put(45,19){\circle*{1.7}}
\put(10,19){\line(35,0){35}}
\put(27.5,21.5){\makebox(0,0){$j_{p}$}}
\put(20,09){\circle*{1.7}}
\put(18,07){\makebox(0,0){$+$}}
\put(35,09){\circle*{1.7}}
\put(37,07){\makebox(0,0){$-$}}
\put(20,09.6){\line(15,0){15}}
\put(20,09.3){\line(15,0){15}}
\put(20,09){\line(15,0){15}}
\put(20,08.7){\line(15,0){15}}
\put(20,08.4){\line(15,0){15}}
\put(26,07){\line(0,4){4}}
\put(30,09.6){\vector(-1,0){3}}
\put(30,08.4){\vector(-1,0){3}}
\put(27.5,13){\makebox(0,0){$J_{2}$}}
\put(9.5,29){\line(1,1){10.5}}
\put(9.4,19){\line(1,-1){5.3}}
\put(9.7,19){\line(1,-1){5.15}}
\put(10,19){\line(1,-1){10}}
\put(10.3,19){\line(1,-1){4.9}}
\put(10.6,19){\line(1,-1){4.75}}
\put(34.4,39){\line(1,-1){5.3}}
\put(34.7,39){\line(1,-1){5.15}}
\put(45,29){\line(-1,1){10}}
\put(35.3,39){\line(1,-1){4.9}}
\put(35.6,39){\line(1,-1){4.75}}
\put(45,19){\line(-1,-1){10}}
\put(10,19){\line(0,10){10}}
\put(7,25.5){\makebox(0,0)[t]{$j_{p}$}}
\put(19.4,24){\line(0,15){15}}
\put(19.7,24){\line(0,15){15}}
\put(20,09){\line(0,30){30}}
\put(20.3,24){\line(0,15){15}}
\put(20.6,24){\line(0,15){15}}
\put(12,13){\makebox(0,0){$y$}}
\put(07.5,29){\makebox(0,0){$\bf{+}$}}
\put(07.5,19){\makebox(0,0){$\bf{+}$}}
\put(44.4,19){\line(0,5){5}}
\put(44.7,19){\line(0,5){5}}
\put(45,19){\line(0,10){10}}
\put(48,25.5){\makebox(0,0)[t]{$x$}}
\put(45.3,19){\line(0,5){5}}
\put(45.6,19){\line(0,5){5}}
\put(35,09){\line(0,30){30}}
\put(43,13){\makebox(0,0){$j_{n}$}}
\put(47.5,29){\makebox(0,0){$\bf{-}$}}
\put(47.5,19){\makebox(0,0){$\bf{+}$}}
\put(27.5,03){\makebox(0,0){$ B_{4}$}}
\end{picture}
\caption{Diagram showing the recoupling coefficient $B_{4}$ of the tensorial product from (\ref{eq:Tensor311}) to (\ref{eq:Tensor312}).}
\label{12j}
\end{center}
\end{figure}
In this case, the transformation of the tensorial product from (\ref{eq:Tensor311}) to (\ref{eq:Tensor312}) results in only one member in (\ref{eq:Tensor312}), where the transformation matrix of the tensorial structure is shown graphically in Fig.~\ref{12j}. $B_4$ is proportional to the coefficient 12$j$-symbol~\cite{Gaigatal:85,JucBan:77a}. The final expression of this rearrangement of the tensorial structure (\ref{eq:Tensor311}) is the following:
\begin{equation}
\label{eq:Tensor313}
\begin{aligned}
\biggl[ \Bigl[ \bigl[ a^{(j_n)} \times \tilde{a}^{(j_m)} \bigr]^{(k)}   \times \bigl[ a^{(j_p)} \times \tilde{a}^{(j_n)} \bigr]^{(x)} \Bigr]^{(k')}  \times \bigl[ a^{(j_m)} \times \tilde{a}^{(j_p)} \bigr]^{(k')} \biggr]^{(0)}
\\
= \;
\sum_{J_1,J_2,y} B_4 \; \biggl[ \Bigl[ \bigl[ \tilde{a}^{(j_m)} \times a^{(j_m)} \bigr]^{(J_1)} \times \bigl[ a^{(j_p)} \times \tilde{a}^{(j_p)} \bigr]^{(y)} \Bigr]^{(J_2)}  \times \bigl[ a^{(j_n)} \times \tilde{a}^{(j_n)} \bigr]^{(J_2)} \biggr]^{(0)}
\end{aligned}
\end{equation}
and
\begin{equation}
\label{eq:Tensor314}
\begin{aligned}
\biggl[ \Bigl[ \bigl[ a^{(j_n)} \times \tilde{a}^{(j_m)} \bigr]^{(k)}   \times \bigl[ a^{(j_p)} \times \tilde{a}^{(j_p)} \bigr]^{(x)} \Bigr]^{(k')}  \times \bigl[ a^{(j_m)} \times \tilde{a}^{(j_n)} \bigr]^{(k')} \biggr]^{(0)}
\\
= \;
\left( -1 \right)^{j_m+j_p+k'+1} \sqrt{\left[ k, k', x \right]} \sum_{J_1,J_2,y} \left( -1 \right)^{J_1+J_2} \sqrt{\left[ J_1, J_2, y \right]}
\\
\times \; \biggl[ \Bigl[ \bigl[ \tilde{a}^{(j_m)} \times a^{(j_m)} \bigr]^{(J_1)} \times \bigl[ a^{(j_p)} \times \tilde{a}^{(j_p)} \bigr]^{(y)} \Bigr]^{(J_2)} \times \bigl[ a^{(j_n)} \times \tilde{a}^{(j_n)} \bigr]^{(J_2)}  \biggr]^{(0)} 
\\
\times \; \sum_{z} \left[ z \right]
		  \left\{
    \begin{array}{ccc}
      J_{1} & j_{n} & z \\
      j_{n} & y     & J_{2}
    \end{array} \right\}
		  \left\{
    \begin{array}{ccc}
      y & j_{n} & z \\
      x & j_{p} & j_{p}
    \end{array} \right\}
		  \left\{
		 \begin{array}{ccc}
      j_{p} & x     & z \\
      k     & j_{m} & k'
    \end{array} \right\}
		  \left\{
		 \begin{array}{ccc}
      j_{m} & k     & z \\
      j_{n} & J_{1} & j_{m}
    \end{array} \right\}.
\end{aligned}
\end{equation}
The program library \texttt{librang}~\cite{Gaigalas:2022} available in the {\sc Grasp} package is sufficient for the calculation of the reduced matrix element of Feynman diagram $A_6$. The use of the library in this case is exactly the same as the one presented in subsection~\ref{sec:PT_SA_direct_Fourth}, which was dealt with in diagram $A_5$. In this case only the multiplier $\Theta ^{\prime }\left( n_i\lambda _i,n_j\lambda _j,n_i^{\prime
}\lambda _i^{\prime },n_j^{\prime }\lambda _j^{\prime },\Xi \right) $ differs.

\subsubsection{The special cases of the fourth type of valence-valence correlations}
\label{sec:PT_SA_special_Fourth}

The way to calculate reduced matrix elements of Feynman diagrams $A_5$ and $A_6$ in the general case is presented in subsections~\ref{sec:PT_SA_direct_Fourth}, and \ref{sec:PT_SA_exchange_Fourth}. But it is possible to extract, as it was shown in subsection~\ref{sec:PT_SA_Third}, the individual parts which are more straightforward to calculate, i.e., the coefficient $\Delta \mathcal{E}_0$ (does not depend on the term) and $\Delta \mathcal{F}^{k}(n,p)$ (regular spin-angular library \texttt{librang} can be used). The extraction of these expressions is the same as in the papers~\cite{Gaigetal:2024CV,Gaigetal:2024C,Gaigetal:2024CC,Gaigetal:2025VV}. Therefore, only the reduced matrix element of part of Feynman diagrams $A_5$ and $A_6$ is calculated from the general expressions, where the rank $k > 0$. The part which is calculated from the general expression will be denoted by $\Delta \widetilde{\mathcal{R}}^{(k, k', x)} \left( m n p  \right)$ in the future.

In the next section, we will give the final expressions for these two types of correlations (third and fourth types of valence-valence correlations), as we did in the papers~\cite{Gaigetal:2024CV,Gaigetal:2024C,Gaigetal:2024CC,Gaigetal:2025VV}.

\section{Combination of relativistic configuration interaction approximation
with the stationary second-order Rayleigh-Schr\"odinger many-body perturbation theory}
\label{Sec:VVImplementation}

Similar to CV, C, CC, and VV correlations \cite{Gaigetal:2024CV,Gaigetal:2024C,Gaigetal:2024CC,Gaigetal:2025VV}, the admixed configurations from VV correlations deriving from three-particle Feynman diagram VV$_3$
(Eqs. (\ref{eq:VVT-a}) and (\ref{eq:VVT-b}))
can be added to usual energy $E_0 \left(K \right)$ of the 
term $\chi J$ of the configuration $K$ and can be
expressed as the energy $\Delta \mathcal{E}_0 \left(K J \right)$, which does not depend on the term, and the sum of the product of Slater integrals and spin-angular coefficients, describing the interaction within and between open subshells:
\begin{eqnarray}
\label{eq:VVBogEnergy}
\hspace*{-2.5cm}
   E\left(K \chi J \right)
	\nonumber \\
& &
   = E_0 \left(K J\right) + \Delta \mathcal{E}_0 \left(K J \right) 
	\nonumber \\  [0.2cm]
& &
	+ \; \sum_{n\ell j} \sum_{k>0} \widetilde{f}_k \left( \ell j^{w}, \; K \chi J  \right)
	\left[ \mathcal{F}^{k} \left( n \ell j, \; n \ell j \right)  
	+ \Delta \mathcal{F}^{k} \left( n \ell j, \; n \ell j \right) \right]
	\nonumber \\
& &
	+ \; \sum_{n\ell j} \sum_{n'\ell'j' > n\ell j} \left\{ \sum_{k>0} \widetilde{f}_k \left( \ell j^{w} \; \ell' j'^{w'},
	\; K \chi J  \right) \right.
\left[ \mathcal{F}^{k} \left( n \ell j, \; n '\ell' j' \right)  
	+ \Delta \mathcal{F}^{k} \left( n \ell j, \; n' \ell' j' \right) \right]
	\nonumber \\ [0.2cm]
& & 
	+ \sum_{k} \widetilde{g}_k \left( \ell j^{w} \; \ell' j'^{w'}, \; K \chi J  \right)	
\mathcal{G}^{k} \left( n \ell j, \; n '\ell' j' \right)  
	\nonumber \\ [0.2cm]
& &
\left.
	+ \sum_{k} \widetilde{v}_k \left( \ell j^{w} \; \ell' j'^{w'}, \ell j^{w-2} \; \ell' j'^{w'+2},
	\; K \chi J \; K' \chi' J \right)	
\mathcal{R}^{k} \left( n \ell j n \ell j, \; n '\ell' j' n '\ell' j' \right) \right\} 
	\nonumber \\
& & 
+ \sum_{\substack{n\ell j \\ n'\ell'j' \, \neq \, n\ell j}}
\; \sum_{\substack{k>0 \\ k',x}} 
\left< \Psi \left\| \biggl[ \bigl[ \tilde{a}^{(j)} \times a^{(j)} \bigr]^{(k)} \times  \Bigl[ \bigl[ a^{(j')} \times \tilde{a}^{(j')} \bigr]^{(x)} \times \bigl[ a^{(j')} \times \tilde{a}^{(j')} \bigr]^{(k')} \Bigr]^{(k)} \biggr]^{(0)} \right\| \Psi \right> 
	\nonumber \\
& & 
\hspace{1.0cm} \times \;
\Delta \widetilde{\mathcal{R}}^{(k,k',x)}
\left( n \ell j \; n '\ell' j' \; n '\ell' j' \right)
	\nonumber \\
& & 
+ \sum_{\substack{n\ell j \\ n'\ell'j' \, \neq \, n\ell j \\ n''\ell''j'' \, \neq \, n\ell j}}
\; \sum_{\substack{k>0 \\ k',x}} 
\left< \Psi \left\| \biggl[ \Bigl[ \bigl[ \tilde{a}^{(j)} \times a^{(j)} \bigr]^{(k)} \times \bigl[ a^{(j')} \times \tilde{a}^{(j')} \bigr]^{(x)} \Bigr]^{(k')} \times \bigl[ a^{(j'')} \times \tilde{a}^{(j'')} \bigr]^{(k')} \biggr]^{(0)} \right\| \Psi \right> 
	\nonumber \\
& & 
\hspace{1.0cm} \times \;
\Delta \widetilde{\mathcal{R}}^{(k,k',x)}
\left( n \ell j \; n '\ell' j' \; n ''\ell'' j'' \right),
\end{eqnarray}
where $\widetilde{f}_k$, $\widetilde{g}_k$, and $\widetilde{v}_k$ are spin-angular coefficients from which submatrix elements $\redmem{\ell j}{\, C^{(k)} \,}{ \ell^{\prime} j^{\prime}}$ are extracted. 
Therefore, summation over $k$ runs over all 
possible values instead of the values that satisfy the triangular condition $\left( \ell \ell^{\prime} k\right)$ as it is in the regular case. The $\mathcal{F}^{k} \left( n \ell j, \; n '\ell' j' \right)$,
$\mathcal{G}^{k} \left( n \ell j, \; n '\ell' j' \right)$, and $\mathcal{R}^{k} \left( n \ell j n \ell j, \; n '\ell' j' n '\ell' j' \right)$ are generalized integrals of electrostatic interaction between electrons. The definition of $\mathcal{R}^{k} \left( n \ell j n \ell j, \; n '\ell' j' n '\ell' j' \right)$ is the following:
\begin{eqnarray}
\label{eq:BogRk}
\hspace*{-2.5cm}
   \mathcal{R}^{k}\left(i j, i' j'\right)
	\nonumber \\
& &
   = \left\{ \left[ 1 + \delta \left( i, j \right) \right]  \left[ 1 + \delta \left( i', j' \right) \right]  \right\} ^{-1/2}
	 \, R^{k}\left(n_i j_i \, n_jj_j, \, n_{i'}j_{i'} \, n_{j'}j_{j'} \right)
\nonumber \\
& &
	   \times \redmem{\ell_i j_{i}}{\, C^{(k)} \,}{ \ell_{i'} j_{i'}}
     \redmem{\ell_j j_{j}}{\, C^{(k)} \,}{ \ell_{j'} j_{j'}}, 
\end{eqnarray}
where $R^{k}\left(n_i j_i \, n_jj_j, \, n_{i'}j_{i'} \, n_{j'}j_{j'} \right)$ is the same radial integral as in 
Eq. (\ref{eq:deffX}). Definitions  $\mathcal{F}^{k} \left( n \ell j, \; n '\ell' j' \right)$,
$\mathcal{G}^{k} \left( n \ell j, \; n '\ell' j' \right)$ straightforwardly follow from Eq. (\ref{eq:BogRk}).
Due to the inherent complexity of the three-particle operator, the expression (\ref{eq:VVBogEnergy}) does not fully distinguish all the members that are
independent of the term. Therefore, a very small number of them are retained in the members $\Delta \widetilde{\mathcal{R}}^{(k,k',x)}\left( n \ell j \; n '\ell' j' \; n '\ell' j' \right)$, and $\Delta \widetilde{\mathcal{R}}^{(k,k',x)}\left( n \ell j \; n '\ell' j' \; n ''\ell'' j'' \right)$.

The contribution deriving from the VV correlations of the configurations $K'$ to $E (K \chi J)$ in the second-order of the perturbation theory
can be extracted from Eq. (\ref{eq:VVBogEnergy}) as
\begin{eqnarray}
\label{eq:BogEnergy_PT}
\hspace*{-2.5cm}
  \Delta E_{PT (VV T)}
	\nonumber \\
& &
   =  \Delta \mathcal{E}_0 \left(K J \right) 
	\nonumber \\  [0.2cm]
& &
	+ \; \sum_{n\ell j} \sum_{k>0} \widetilde{f}_k \left( \ell j^{w}, \; K \chi J  \right)
	\Delta \mathcal{F}^{k} \left( n \ell j, \; n \ell j \right)
	\nonumber \\
& &
	+ \; \sum_{n\ell j} \sum_{n'\ell'j' > n\ell j} \sum_{k>0} \widetilde{f}_k \left( \ell j^{w} \; \ell' j'^{w'},
	\; K \chi J  \right)
 \Delta \mathcal{F}^{k} \left( n \ell j, \; n' \ell' j' \right) 
	\nonumber \\ [0.2cm]
& &
+ \sum_{\substack{n\ell j \\ n'\ell'j' \, \neq \, n\ell j}}
\; \sum_{\substack{k>0 \\ k',x}} 
\left< \Psi \left\| \biggl[ \bigl[ \tilde{a}^{(j)} \times a^{(j)} \bigr]^{(k)} \times  \Bigl[ \bigl[ a^{(j')} \times \tilde{a}^{(j')} \bigr]^{(x)} \times \bigl[ a^{(j')} \times \tilde{a}^{(j')} \bigr]^{(k')} \Bigr]^{(k)} \biggr]^{(0)} \right\| \Psi \right> 
	\nonumber \\
& & 
\hspace{1.0cm} \times \;
\Delta \widetilde{\mathcal{R}}^{(k,k',x)}
\left( n \ell j \; n '\ell' j' \; n '\ell' j' \right)
	\nonumber \\
& & 
+ \sum_{\substack{n\ell j \\ n'\ell'j' \, \neq \, n\ell j \\ n''\ell''j'' \, \neq \, n\ell j}}
\; \sum_{\substack{k>0 \\ k',x}} 
\left< \Psi \left\| \biggl[ \Bigl[ \bigl[ \tilde{a}^{(j)} \times a^{(j)} \bigr]^{(k)} \times \bigl[ a^{(j')} \times \tilde{a}^{(j')} \bigr]^{(x)} \Bigr]^{(k')} \times \bigl[ a^{(j'')} \times \tilde{a}^{(j'')} \bigr]^{(k')} \biggr]^{(0)} \right\| \Psi \right> 
	\nonumber \\
& & 
\hspace{1.0cm} \times \;
\Delta \widetilde{\mathcal{R}}^{(k,k',x)}
\left( n \ell j \; n '\ell' j' \; n ''\ell'' j'' \right).
\end{eqnarray}

\begin{table*}
\begin{center}
\begin{tabular}{|l|} \hline
$\Delta \mathcal{E}_0$ corrections \\ \hline \hline
\\
$\overbrace{(n_{m} \ell_{m})\, j_{m}^{w_m} \; (n_{n} \ell_{n})\, j_{n}^{w_n}}^{\text{valence subshells}} \;
\rightarrow \; \overbrace{(n_{m} \ell_{m})\, j_{m}^{w_m+1} \; (n_{n} \ell_{n})\, j_{n}^{w_n-2}}^{\text{valence subshells}} \; \overbrace{(n_{r} \ell_{r})\, j_{r}}^{\text{virtual subshell}}$ \\
\\
$\underbrace{- \frac{2 \left( \left[ j_m \right] - w_m \right) w_n \left(w_n -1 \right)}{\left[ j_m, j_n\right]} \; \mathcal{A'}\left( 0, \, n n, \, m r \right)
	}_{\text{from $\text{VV}_3$ Feynman diagram}}$ \\
\\ \hline \hline
\\
$\overbrace{(n_{m} \ell_{m})\, j_{m}^{w_m} \; (n_{n} \ell_{n})\, j_{n}^{w_n} \; (n_{p} \ell_{p})\, j_{p}^{w_p}}^{\text{valence subshells}} \;
\rightarrow \; \overbrace{(n_{m} \ell_{m})\, j_{m}^{w_m+1} \; (n_{n} \ell_{n})\, j_{n}^{w_n-1} \; (n_{p} \ell_{p})\, j_{p}^{w_p-1}}^{\text{valence subshells}} \; \overbrace{(n_{r} \ell_{r})\, j_{r}}^{\text{virtual subshell}}$ \\
\\ $\underbrace{- \left( -1 \right)^{j_m+j_n+j_p+j_r}\frac{\left( \left[ j_m \right] - w_m \right) w_n w_p}{\left[ j_m, j_n, j_p \right]} \; \sum_{k}  \left\{ \frac{ \mathcal{P}\left( kk, \, n p, \, m r \right)}{\left[ k\right]} + \mathcal{C}\left( k, \, n p, \, m r \right) \right\} \biggl( 1 + \mbox{P}\left( n \rightleftharpoons p \right) \biggr)
	}_{\text{from $\text{VV}_3$ Feynman diagram}}$ \\
\\ \hline 
\end{tabular}
\end{center}
\caption{Expressions for valence-valence corrections to the energy in Eqs. (\ref{eq:VVBogEnergy}) and (\ref{eq:BogEnergy_PT}), independent of the term.}
\label{tab:Implemen_VV1}
\end{table*}

\begin{table*}
\begin{center}
\begin{tabular}{|lcc|} \hline
Corrections & Slater integral & $k$ values\\ \hline  \hline
& & \\
\multicolumn{3}{|c|}{$\overbrace{(n_{m} \ell_{m})\, j_{m}^{w_m} \; (n_{n} \ell_{n})\, j_{n}^{w_n}}^{\text{valence subshells}} \;
\rightarrow \; \overbrace{(n_{m} \ell_{m})\, j_{m}^{w_m+1} \; (n_{n} \ell_{n})\, j_{n}^{w_n-2}}^{\text{valence subshells}} \; \overbrace{(n_{r} \ell_{r})\, j_r}^{\text{virtual subshell}}$} \\
& &\\
$\underbrace{ 
-4 \left[ k \right] \frac{\left( \left[ j_m \right] - w_m \right)}{\left[ j_m \right]}  \mathcal{A^{\prime}}\left( k, \, n n, \, m r \right) }_{\text{from $\text{VV}_{3}$ Feynman diagram}}$
&$\Delta \mathcal{F}^{k}(n,n)$ & $k>0$\\
& &\\
& &\\
$\underbrace{ 
2 \left[ k \right] \sum_{k'} \left( -1 \right)^{j_n+j_r+k'}
		  \left\{
    \begin{array}{ccc}
      j_{m} & j_{m} & k \\
      j_{n} & j_{n} & k'
    \end{array} \right\}
		\mathcal{C}\left( k', \, r m, \, n n \right)}_{\text{from $\text{VV}_{3}$ Feynman diagram}}$ 
&$\Delta \mathcal{F}^{k}(m,n)$ & $k>0$\\
& &\\
& &\\
$\underbrace{ 
2 
\left( -1 \right)^{-j_m + j_n + x} \sqrt{\left[ k, k', x \right]} \; \mathcal{G}\left( k \, k' \, x, \, n n, \, m r \right)}_{\text{from $\text{VV}_{3}$ Feynman diagram}}$
&$\Delta \widetilde{\mathcal{R}}^{(k, k', x)} \left( m n n \right)$ & $k>0$ \\
& & \\ \hline \hline
\end{tabular}
\end{center}
\caption{Expressions for Slater integrals $\Delta \mathcal{F}^{k}(n,n)$, $\Delta \mathcal{F}^{k}(m,n)$, and $\Delta \widetilde{\mathcal{R}}^{(k, k', x)} \left( m n n  \right)$
(see Eqs. (\ref{eq:VVBogEnergy}) and (\ref{eq:BogEnergy_PT})) corrections corresponding 
to the third type of valence-valence correlations.} 
\label{tab:Implemen_VV2}
\end{table*}

\begin{table*}
\begin{center}
\begin{tabular}{|lcc|} \hline
Corrections & Slater integral & $k$ values\\ \hline  \hline
& & \\
\multicolumn{3}{|c|}{$\overbrace{(n_{m} \ell_{m})\, j_{m}^{w_m} \; (n_{n} \ell_{n})\, j_{n}^{w_n} \; (n_{p} \ell_{p})\, j_{p}^{w_p}}^{\text{valence subshells}} \;
\rightarrow \; \overbrace{(n_{m} \ell_{m})\, j_{m}^{w_m+1} \; (n_{n} \ell_{n})\, j_{n}^{w_n-1} \; (n_{p} \ell_{p})\, j_{p}^{w_p-1}}^{\text{valence subshells}} \; \overbrace{(n_{r} \ell_{r})\, j_{r}}^{\text{virtual subshell}}$} \\
& &\\
$\underbrace{ 
 \left( -1 \right)^{j_m+j_n} \, \left[k\right]\sqrt{\left[ k \right]} \, \frac{\left( \left[ j_m \right] - w_m \right)}{\sqrt{\left[j_m\right]}} \; \Biggl\{ \left( -1 \right)^{k+1}  \; \mathcal{G}\left( 0 \, k \, k, \, n p, \, m r \right) \Biggr.}_{\text{from $\text{VV}_{3}$ Feynman diagram}}$
&$\Delta \mathcal{F}^{k}(n,p)$ & $k>0$\\
& &\\
$\underbrace{ \Biggl.+ 
 \sum_{k_1, k_2, k_3}
 \left( -1 \right) ^{k_1+k} \left[ k_3 \right] \;
		  \left\{
    \begin{array}{ccc}
      j_{n} & j_{p} & k_3 \\
      k_{1} & k_{2} & j_r
    \end{array} \right\}
\mathcal{Q} \left( k_1k_2, n p, m r \right) }_{\text{from $\text{VV}_{3}$ Feynman diagram} }$ & & \\
& & \\
\hspace{0.5cm} $\underbrace{ \Biggl. \times \;
\mathcal{C}_{12j}\left( j_m j_n j_p, \, k_1 \, k_2 \, k_3, \, 0 \, k \, k \right) \Biggr\} \;
\Biggl( 1 + \mbox{P} 
		  \left( \hspace{-0.15cm}
\begin{array}{lcl}
      n    &\hspace{-0.25cm}\rightleftharpoons&\hspace{-0.25cm}p \\
      k_{1}&\hspace{-0.25cm}\rightleftharpoons&\hspace{-0.25cm}k_{2}
    \end{array}  \hspace{-0.15cm} \right)
 \Biggr) }_{\text{from $\text{VV}_{3}$ Feynman diagram} }$
& &\\
& &\\
& &\\
$\underbrace{ 
\left( -1 \right)^{j_m + j_n} \sqrt{\left[ k, k^{'}, x \right]} \; \Biggl\{ \left( -1 \right)^{x+1}
\mathcal{G}\left( k \, k^{'} \, x, \, n p, \, m r \right) \Biggr.}_{\text{from $\text{VV}_{3}$ Feynman diagram}}$
&$\Delta \widetilde{\mathcal{R}}^{(k, k^{'}, x)} \left( m n p \right)$ & $k>0$ \\
& & \\
\hspace{0.5cm} $\underbrace{ 
+ \sum_{k_1, k_2, k_3} \left( -1 \right)^{k + k^{'} + k_1} \left[ k_3 \right]
		  \left\{
    \begin{array}{ccc}
      j_{n} & j_{p} & k_3 \\
      k_{1} & k_{2} & j_r
    \end{array} \right\}
 \mathcal{Q} \left( k_1k_2, n p, m r \right)
		}_{\text{from $\text{VV}_{3}$ Feynman diagram} }$ & & \\
& &\\
\hspace{0.5cm} $\underbrace{ \times 
\; \mathcal{C}_{12j}\left( j_m j_n j_p, \, k_1 \, k_2 \, k_3, \, k \, k^{'} \, x \right)\Biggl\} 
 \;
\Biggl( 1 + \mbox{P} 
		  \left( \hspace{-0.15cm}
\begin{array}{lcl}
      n    &\hspace{-0.25cm}\rightleftharpoons&\hspace{-0.25cm}p \\
      k_{1}&\hspace{-0.25cm}\rightleftharpoons&\hspace{-0.25cm}k_{2} \\
			k' &\hspace{-0.25cm}\rightleftharpoons&\hspace{-0.25cm}x 
    \end{array}  \hspace{-0.15cm} \right)
 \Biggr)}_{\text{from $\text{VV}_{3}$ Feynman diagram}}$ & & \\
& &\\ \hline 
\end{tabular}
\end{center}
\caption{Expressions for Slater integral $\Delta \mathcal{F}^{k}(n,p)$ and $\Delta \widetilde{\mathcal{R}}^{(k, k', x)} \left( m n p  \right)$ 
(see Eqs. (\ref{eq:VVBogEnergy}) and (\ref{eq:BogEnergy_PT})) corrections corresponding 
to the fourth type of valence-valence correlations.} 
\label{tab:Implemen_VV3}
\end{table*}

The contribution of the third and fourth type of the VV correlations in the second-order of the perturbation theory is expressed over $\Delta \mathcal{E}_0 \left(K J \right)$, $\Delta \mathcal{F}^{k} \left( n \ell j, \; n \ell j \right)$, $\Delta \mathcal{F}^{k} \left( n \ell j, \; n' \ell' j' \right)$, 
$\Delta \widetilde{\mathcal{R}}^{(k,k',x)}\left( n \ell j \; n '\ell' j' \; n '\ell' j' \right)$, and $\Delta \widetilde{\mathcal{R}}^{(k,k',x)}
\left( n \ell j \; n '\ell' j' \; n ''\ell'' j'' \right)$ (see 
Table~\ref{tab:Implemen_VV1},  \ref{tab:Implemen_VV2}, and \ref{tab:Implemen_VV3}).
These formulae are additionally expressed via the quantities

\begin{equation}
\label{eq:BogAp}
   \mathcal{A^{\prime}}\left(x, \; i j, \; i' j'\right) 
  = \sum_{k,k'}
		  \left\{
    \begin{array}{ccc}
      k  & k' & x \\
      j_{i} & j_{i} & j_{i'}
    \end{array} \right\}
				  \left\{
    \begin{array}{ccc}
      k  & k' & x \\
      j_{j} & j_{j} & j_{j'}
    \end{array} \right\}
\mathcal{P}\left(kk', \; i j, \; i' j'\right) ,
\end{equation}
\begin{equation}
\label{eq:BogC}
   \mathcal{C}\left(k, \; i j, \; i' j'\right) 
  = \sum_{k'}
		  \left\{
    \begin{array}{ccc}
      k  & j_{i} & j_{i'} \\
      k' & j_{j} & j_{j'}
    \end{array} \right\}
\mathcal{Q}\left(kk', \; i j, \; i' j'\right) ,
\end{equation}
and
\begin{equation}
\label{eq:BogG}
   \mathcal{G}\left( x_1 \, x_2 \, x, \; i j, \; i' j'\right) 
  = \sum_{k,k'}
	   \left( -1 \right)^{k'}
		  \left\{
    \begin{array}{ccc}
      j_{j} & j_{j} & x \\
      k     & k'    & j_{j'}
    \end{array} \right\}
				  \left\{
    \begin{array}{ccc}
		  j_{i'} & j_{i} & k \\
      x_1    & x_2   & x \\
      j_{i'} & j_{i} & k'
    \end{array} \right\}
\mathcal{P}\left(kk', \; i j, \; i' j'\right) ,
\end{equation}
where
\begin{equation}
\label{eq:BogP}
   \mathcal{P}\left(kk', \; i j, \; i' j'\right) 
   = \mathcal{R}^{k}\left(i j, \; i' j'\right) \; \mathcal{R}^{k'}\left(i' j', \; i j \right)  \; 
	\mathcal{O}\left(K', K \right) ,
\end{equation}

\begin{equation}
\label{eq:BogQ}
   \mathcal{Q}\left(kk', \; i j, \; i' j'\right)
   = \mathcal{R}^{k}\left(i j, \; i' j'\right) \; \mathcal{R}^{k'}\left(i' j', \; j i\right)  \; 
\mathcal{O}\left(K', K \right) ,
\end{equation}

\begin{equation}
\label{eq:BogO1}
\mathcal{O}\left(K', K \right)
= \frac{1}{\overline{E}\left(K' \right)-\overline{E}\left(K\right)} ,
\end{equation}
where $\overline{E}\left(K\right)$ is the averaged energy of the 
state 
for which calculations are performed. $\overline{E}\left(K^{'}\right)$ is the averaged energy for the admixed configuration $K'$.
For details on how to find $\overline{E}\left(K\right)$ and $\overline{E}\left(K^{'}\right)$, see \cite[Section~3]{{Gaigetal:2024CV}}.
We would like to emphasize that the energy denominator (\ref{eq:BogO1}) is defined differently/opposite to the expressions of Feynman diagrams 
(see Fig.~\ref{VV_3}, Eqs. (\ref{eq:BogP}), (\ref{eq:BogQ})).

\begin{eqnarray}
\label{eq:BogC12j}
\hspace*{-0.7cm}
   \mathcal{C}_{12j}\left( i j i', \; k_1 k_2 k_3, \; J_1 J_2 J \right)
\nonumber \\
& & 
\hspace*{-3.2cm}
  = \sum_{x}
	   \left[ x \right]
		  \left\{
    \begin{array}{ccc}
      J_{1} & j_{j} & x \\
      j_{j} & J     & J_{2}
    \end{array} \right\}
		  \left\{
    \begin{array}{ccc}
      J     & j_{j}  & x \\
      k_{3} & j_{i'} & j_{i'}
    \end{array} \right\}
		  \left\{
		 \begin{array}{ccc}
      j_{i'} & k_{3} & x \\
      k_{1}  & j_{i} & k_{2}
    \end{array} \right\}
		  \left\{
		 \begin{array}{ccc}
      j_{i} & k_{1} & x \\
      j_{j} & J_{1} & j_{i}
    \end{array} \right\} .
\end{eqnarray}

We would also like to point out that the notation of ranks such as $J_1$, $J_2$, $y$ in the multipliers of this section under $\Delta \widetilde{\mathcal{R}}^{(k, k', x)} \left( m n n  \right)$ and $\Delta \widetilde{\mathcal{R}}^{(k, k', x)} \left( m n p  \right)$ have been renamed for convenience to make the expressions in Table \ref{tab:Implemen_VV2} and \ref{tab:Implemen_VV3} more straightforward and more understandable.

This theory in irreducible tensorial form, as in the papers~\cite{Gaigetal:2024CV,Gaigetal:2024C,Gaigetal:2024CC,Gaigetal:2025VV}, is more suitable to be included in such version of the {\sc Grasp} which is based on configuration state function generators~\cite{grasp2023,grasp2025}. This is related to the fact that this version of the software package allows us to easily distinguish $F$, $F'$, and $G$ sets of orbitals in the process of computing atomic data. In the following section, we will present a test case of this implementation.

\section{Calculation of core-valence, core and valence-valence 
(including these which are described by the three-particle Feynman diagram) correlations with a new approach}
\label{Sec:Calculations}
The computations in the present work were performed in the regular way and using the method
based on the Rayleigh-Schr\"odinger perturbation theory 
in an irreducible tensorial form \cite{Gaigetal:2024CV,Gaigetal:2024C,Gaigetal:2024CC,Gaigetal:2025VV} 
which was extended to include VV correlations described by the three-particle Feynman diagram (VVT).
The developed RSMBPT method was applied to select the most significant configuration state functions. 
For the first time in this work, the configuration state function (CSF) bases constructed using the RSMBPT method were used 
to solve the self-consistent field equations.
It should be mention that in previous studies 
\cite{Gaigetal:2024CV,Gaigetal:2024C,Gaigetal:2024CC,Gaigetal:2025VV,Gaigalas_Ce_2024}, 
the RSMBPT method was only applied with relativistic configuration interaction (RCI).
In the case of the regular calculations to include different types of correlations 
in the multiconfiguration Dirac-Hartree-Fock (MCDHF) calculations
often is a complex task, or even impossible, due to time and computer resource limitations, especially for complex systems.
Basically, only VV correlations and additionally some limited correlations from core (if the CSF bases are not to large) 
are included in these calculations.
The application of the RSMBPT method for solving self-consistent field equations
allows to include the most important correlations of various types.

For such investigations, the energy structure calculations were performed for  
105 lowest energy levels of the $\mathrm{4s^24p^2}$, $\mathrm{4p^4}$, 
$\mathrm{4s^24p\{4d,4f,5s,5p,5d,6s,6p\}}$, $\mathrm{4s4p^3}$, and 
$\mathrm{4s4p^2\{4d,5s\}}$ configurations of the Se~III using the regular way and the RSMBPT method 
when CV, C, VV and VVT correlations were included. In further description, VV correlations investigated in Ref.
\cite{Gaigetal:2025VV} and the VV correlations described by the three-particle Feynman diagram (VVT) investigated in this work
would be marked as VV.
The multireference (MR) set in the present calculations consists of the $\mathrm{4s^24p^2}$, $\mathrm{4p^4}$, 
$\mathrm{4s^24p\{4f,5p,6p\}}$, $\mathrm{4s4p^2\{4d,5s\}}$ even and 
$\mathrm{4s^24p\{4d,5s,5d,6s\}}$, $\mathrm{4s4p^3}$ odd configurations. 
Firstly, the MCDHF calculation for 
even and odd states of the configurations belonging to MR set was done in the extended optimal level (EOL) scheme \citep{EOL}.
These radial wavefunctions were used for further investigations.
The initial calculation was followed by
separate calculations in the EOL scheme for the even and odd parity states using the regular way and the RSMBPT method. 
These calculations are described in the subsections below, and the results are presented in section \ref{Results}.
The calculations using both methods were performed including only CSFs that have non-zero
matrix elements in the sets of spin-angular integration with the CSFs belonging to the configurations
in the MR.
At the RCI calculations step the 
Breit interactions and leading quantum electrodynamic effects – the vacuum polarization and
self-energy corrections were included.

\subsection{Computational schemes}
\label{Computational_schemes}
\subsubsection{Regular {\sc Grasp}2018 calculations}
Regular MCDHF computations including CV, C and VV correlations are marked as \textbf{CV+C+VV MCDHF}.
In this computational scheme single-double (SD) substitutions are allowed from 
the 4s, $\mathrm{4p_-}$, 4p, $\mathrm{4d_-}$, 4d, $\mathrm{4f_-}$, 4f,
5s, $\mathrm{5p_-}$, 5p, $\mathrm{5d_-}$, 5d, 6s, $\mathrm{6p_-}$, 6p valence orbitals of the MR set
and S substitutions from the 3s, $\mathrm{3p_-}$, 3p, $\mathrm{3d_-}$ and 3d core orbitals
to orbital set (OS) $OS_1$=\{7s,$\mathrm{7p_-}$,7p,$\mathrm{6d_-}$,6d,$\mathrm{5f_-}$,5f,$\mathrm{5g_-}$,5g\}, 
$OS_2$=\{8s,$\mathrm{8p_-}$,8p,$\mathrm{7d_-}$,7d,$\mathrm{6f_-}$,6f,$\mathrm{6g_-}$,6g\}.
The 1s, 2s, $\mathrm{2p_-}$ and 2p subshells are defined as inactive core subshells.
The radial wavefunctions of the new OS are estimated using the Thomas-Fermi potential and further self-consistent field equations are solved.
When a new OS is computed, the previous orbitals are frozen. 
Based on the orbitals from the MCDHF calculations, further 
RCI calculations are performed. Regular RCI calculations are marked as \textbf{CV+C+VV RCI}.

\subsubsection{Calculations using the RSMBPT method}
The CSF space in the computations using the RSMBPT method is divided into three sets: $F$, $F'$ and $G$ 
(see Ref. \cite{Gaigetal:2024CV} for details). 
The 1s, 2s, $\mathrm{2p_-}$ and 2p subshells are defined as inactive core subshells in the calculations, the same 
as it is done in the regular {\sc Grasp}2018 calculations. 
The 3s, $\mathrm{3p_-}$, 3p, $\mathrm{3d_-}$ and 3d subshells are defined as core subshells 
(that correspond to $F$ set), 
4s, $\mathrm{4p_-}$, 4p, $\mathrm{4d_-}$, 4d, $\mathrm{4f_-}$, 4f,
5s, $\mathrm{5p_-}$, 5p, $\mathrm{5d_-}$, 5d, 6s, $\mathrm{6p_-}$, 6p as valence subshells 
(that correspond to $F'$ set), and subshells belonging to $OS_1$ and $OS_2$ as virtual ones 
(that correspond to $G$ set). 
Such space distribution is consistent with regular {\sc Grasp}2018 calculations described above.

The RSMBPT calculation procedure is analogous to that used in previous research \cite{Gaigetal:2024CV,Gaigetal:2024C,Gaigetal:2024CC,Gaigetal:2025VV}. 
The contribution of each $K'$ configuration for CSF 
for which energy needs to be calculated according to Rayleigh-Schr\"odinger perturbation theory 
in an irreducible tensorial form is computed according 
to the Eq. (22) of Ref. \cite{Gaigetal:2024CV} (for CV correlations), Eq. (6) of Ref. \cite{Gaigetal:2024C} (for C correlations),
Eq. (19) of Ref. \cite{Gaigetal:2025VV} (for VV correlations) and Eq. (\ref{eq:BogEnergy_PT}) (for VVT correlations).
$K'$ configurations are sorted in descending order according to the impact of the correlations for each level.
Further, $K'$ configurations are selected by CV, C and VV correlations impact
with the specified fraction (expressed in the percentage: 95, 99, 99.5, 99.95 and 100\%) 
of the total correlations contribution.
It should be noted that the program gives the contribution of the correlations of $K'$ configuration 
with a value greater than {\tt 1.0E-11}. Contributions of smaller magnitudes are neglected.
The estimation of correlations using the stationary second-order Rayleigh–Schrödinger many-body perturbation theory 
in an irreducible tensorial form is done for the Coulomb interaction.
The C correlations (Eq. (3) of Ref. \cite{Gaigetal:2024C}) and CV correlations (Eq. (4) of Ref. \cite{Gaigetal:2024C}) 
which are not included with RSMBPT method, were added to calculations in a regular way.

Firstly, the radial wavefunctions of the $OS_1$ are estimated using the Thomas-Fermi potential. 
Then using the RSMBPT method the CSF basis is constructed by selecting the most important CV, C and VV correlations 
with the specified fraction (95, 99, 99.5, 99.95 and 100\%) for even and odd parity.
The self-consistent field equations are solved with the CSF basis for the specified fraction.
The computed radial wavefunctions are taken as initial and the selection procedure 
of the most significant CV, C and VV correlations using the RSMBPT method is performed for construction of CSF basis.
The variation and selection of CSFs is repeated few times to reach the convergence. 
The results of these investigations will be presented in the next section.
Further the radial wavefunctions of the $OS_2$ are computed. In this step firstly 
the radial wavefunctions of the $OS_1$ and $OS_2$ are estimated using the Thomas-Fermi potential 
and selection procedure of the most important correlations from $OS_1$ and $OS_2$ is performed.
This is done so that the contribution of CSFs 
from both ($OS_1$ and $OS_2$) sets would be estimated with similar accuracy radial wavefunctions.
The constructed CSF basis is used to solve self-consistent field equations for $OS_2$, in which 
the radial wavefunctions of the $OS_1$ is taken from final $OS_1$ calculations and are frozen.
The computed radial wavefunctions are taken as initial and the selection procedure of CSFs followed by 
a solution of the self-consistent field equations is repeated a few times to achieve the convergence, as it was done for $OS_1$. 
Results from MCDHF computations including CV, C and VV correlations according to the RSMBPT method are marked as
\textbf{CV+C+VV~MCDHF~(RSMBPT)}.

RCI computations are performed for the $OS_1$ and $OS_2$ taking the radial wavefunctions 
from the specified fraction (95, 99, 99.5, 99.95 and 100\%) MCDHF calculations and CSFs basis from 
the regular {\sc Grasp}2018 calculations.
Additionally, for the $OS_2$ RCI are performed taking the radial wavefunctions 
from the specified fraction (95, 99, 99.5 and 99.95\%) MCDHF calculations and CSFs basis constructed using the RSMBPT method
with the same specified fraction (95, 99, 99.5 and 99.95\%) as in MCDHF.
Results including CV, C and VV correlations according to the RSMBPT method are marked as
\textbf{CV+C+VV~RCI~(RSMBPT)}.

\subsection{Results}
\label{Results}
\subsubsection{Results from MCDHF calculations}
This section presents the results of the investigation of the RSMBPT method 
for solving self-consistent field equations.
These results are also compared with the regular {\sc Grasp}2018 calculations.
In Table \ref{even_1L_selfconsistency} the {\sc self-consistency} and {\sc norm-1} parameters \cite{Jonsetal_grasp_manual} from the MCDHF equations solutions
for the $OS_1$ of the even parity states in cases 95, 99, 99.5, 99.95, 100\% and regular are presented. 
For each step of variation the initial and final results of these parameters are given. 
As seen from Table \ref{even_1L_selfconsistency}, these parameters converge with each variation step, and after a few variations
the initial and final results almost do not change in the case 100\%.
In other cases 95, 99, 99.5, 99.95\%  the convergence is slower, so more variation steps are needed.
The final {\sc self-consistency} and {\sc norm-1} parameters are also very good when the most important CV+C+VV correlations with the 
specified fraction (95, 99, 99.5, 99.95\%) are included in the computations.
In these cases, the CSF bases decreases by up to several times compared to regular calculations.
Similar trends are observed for $OS_1$ of odd parity, also for $OS_2$ of both even and odd parity 
therefore only results for boundary computed cases are presented in Tables \ref{odd_1L_selfconsistency}-\ref{odd_2L_selfconsistency}.

{\footnotesize
\begin{longtable}{l r r r r r r r r r r r}
\caption{\label{even_1L_selfconsistency} {\sc self-consistency} and {\sc norm-1} parameters solving the MCDHF equations of the $OS_1$ for even parity states in cases 95, 99, 99.5, 99.95, 100\% and regular.
Columns with 'x' var. means the number of variations with constructed CSF basis. 
In. and Fin. means the initial and final results of these parameters solving the MCDHF equations.}\\
\hline\hline
\multicolumn{1}{c}{Subshell}& \multicolumn{2}{c}{1 var.} && \multicolumn{2}{c}{2 var.} && \multicolumn{2}{c}{3 var.} && \multicolumn{2}{c}{4 var.}  \\
\cline{2-3} \cline{5-6} \cline{8-9} \cline{11-12}
& \multicolumn{1}{c}{In.} & \multicolumn{1}{c}{Fin.} && \multicolumn{1}{c}{In.} & \multicolumn{1}{c}{Fin.} && \multicolumn{1}{c}{In.} & \multicolumn{1}{c}{Fin.} && \multicolumn{1}{c}{In.} & \multicolumn{1}{c}{Fin.}   \\
\hline
\noalign{\smallskip}
\endfirsthead
\caption{Continued.}\\
\hline\hline
\multicolumn{1}{c}{Subshell}& \multicolumn{2}{c}{1 var.} && \multicolumn{2}{c}{2 var.} && \multicolumn{2}{c}{3 var.} && \multicolumn{2}{c}{4 var.}  \\
\cline{2-3} \cline{5-6} \cline{8-9} \cline{11-12}
& \multicolumn{1}{c}{In.} & \multicolumn{1}{c}{Fin.} && \multicolumn{1}{c}{In.} & \multicolumn{1}{c}{Fin.} && \multicolumn{1}{c}{In.} & \multicolumn{1}{c}{Fin.} && \multicolumn{1}{c}{In.} & \multicolumn{1}{c}{Fin.}   \\
\hline
\noalign{\smallskip}
\endhead
\noalign{\smallskip}
\hline
\hline
\endfoot
\noalign{\smallskip}
\multicolumn{12}{c}{{\sc self-consistency} in case 95\%} \\
7s & 1.81E-02& 2.64E-06&& 2.28E-04& 4.33E-07&& 5.38E-06& 3.79E-07&& 1.67E-06& 8.78E-08 \\
$\mathrm{7p_-}$& 1.15E-02& 3.14E-06&& 1.27E-03& 3.15E-07&& 4.69E-05& 3.75E-07&& 6.12E-06& 1.63E-07 \\
7p & 1.90E-02& 3.92E-06&& 1.00E-03& 4.43E-07&& 6.31E-05& 5.47E-07&& 4.24E-06& 1.70E-07 \\
$\mathrm{6d_-}$& 3.98E-02& 3.11E-06&& 8.25E-04& 3.94E-07&& 3.24E-05& 3.95E-07&& 1.54E-06& 1.31E-07 \\
6d & 5.43E-02& 3.86E-06&& 7.64E-04& 4.15E-07&& 5.49E-05& 4.54E-07&& 5.82E-06& 1.40E-07 \\
$\mathrm{5f_-}$& 2.10E-02& 1.02E-06&& 1.14E-03& 9.71E-08&& 5.72E-05& 1.34E-07&& 1.26E-05& 1.02E-07 \\
5f & 2.66E-02& 1.33E-06&& 1.03E-03& 1.10E-07&& 8.63E-05& 1.95E-07&& 1.26E-05& 9.47E-08 \\
$\mathrm{5g_-}$& 8.45E-03& 1.15E-07&& 4.24E-04& 1.90E-08&& 5.72E-05& 1.45E-07&& 7.87E-07& 8.94E-09 \\
5g & 9.43E-03& 1.24E-07&& 4.46E-04& 1.33E-08&& 3.22E-05& 4.11E-08&& 3.22E-08& 1.28E-08 \\
\multicolumn{12}{c}{{\sc norm-1} in case 95\%} \\                                                     
7s & 3.78E-01& -3.58E-05&& -2.72E-03& -5.36E-06&&  6.35E-05& -4.69E-06&&  1.76E-05& -1.12E-06 \\
$\mathrm{7p_-}$& 1.44E-01& -2.25E-05&& -9.27E-03& -3.82E-06&& -4.61E-04& -4.38E-06&& -6.77E-05& -1.81E-06 \\
7p & 1.57E-01& -2.86E-05&& -6.43E-03& -3.47E-06&& -4.94E-04& -4.58E-06&& -3.22E-05& -1.41E-06 \\
$\mathrm{6d_-}$& 1.35E-01& -1.60E-05&& -4.55E-03& -2.47E-06&& -7.01E-06& -2.25E-06&&  4.48E-06& -7.90E-07 \\
6d & 1.67E-01& -1.70E-05&& -2.03E-03& -1.99E-06&& -1.25E-05& -2.05E-06&&  3.47E-06& -6.43E-07 \\
$\mathrm{5f_-}$& 8.54E-02& -8.77E-06&& -7.73E-03& -8.07E-07&& -5.17E-04& -1.10E-06&& -9.83E-05& -8.04E-07 \\
5f & 1.03E-01& -9.51E-06&& -5.84E-03& -7.81E-07&& -6.61E-04& -1.37E-06&& -8.08E-05& -6.57E-07 \\
$\mathrm{5g_-}$& 4.80E-01&  3.68E-06&& -1.28E-02&  6.16E-07&& -2.00E-03& -5.38E-06&& -2.47E-05&  2.81E-07 \\
5g & 4.81E-01&  3.45E-06&& -1.20E-02&  3.69E-07&& -8.74E-04& -1.38E-06&& -7.48E-07&  3.76E-07 \\ \hline
\multicolumn{12}{c}{{\sc self-consistency} in case 99\%} \\
7s & 1.92E-02& 7.24E-07&& 1.64E-04& 9.53E-07&& 9.41E-06& 1.35E-07&& 7.47E-07& 4.56E-08 \\
$\mathrm{7p_-}$& 1.24E-02& 6.64E-07&& 2.68E-04& 5.96E-07&& 5.82E-05& 7.63E-08&& 2.82E-06& 1.17E-08 \\
7p & 2.11E-02& 8.21E-07&& 3.03E-04& 8.41E-07&& 9.31E-05& 1.05E-07&& 9.87E-07& 2.15E-08 \\
$\mathrm{6d_-}$& 2.56E-02& 6.85E-07&& 2.47E-04& 3.81E-06&& 2.96E-04& 1.66E-07&& 1.86E-06& 9.85E-08 \\
6d & 4.41E-02& 8.19E-07&& 2.41E-04& 1.33E-06&& 2.06E-04& 1.70E-07&& 1.53E-06& 9.87E-08 \\
$\mathrm{5f_-}$& 2.24E-02& 2.02E-07&& 3.59E-04& 1.62E-06&& 4.90E-04& 5.25E-08&& 7.38E-06& 2.95E-08 \\
5f & 2.78E-02& 2.48E-07&& 1.45E-04& 9.31E-07&& 3.17E-04& 4.95E-08&& 3.34E-06& 3.02E-08 \\
$\mathrm{5g_-}$& 8.80E-03& 4.91E-08&& 2.15E-03& 1.20E-05&& 1.97E-03& 5.89E-08&& 3.80E-05& 8.28E-08 \\
5g & 9.88E-03& 1.37E-07&& 3.60E-03& 1.22E-06&& 1.49E-03& 4.08E-08&& 2.87E-05& 6.17E-08 \\
\multicolumn{12}{c}{{\sc norm-1} in case 99\%} \\
7s & 3.91E-01& -8.96E-06&& -2.02E-03& -1.13E-05&&  7.13E-05& -1.64E-06&& -7.87E-06& -5.79E-07 \\
$\mathrm{7p_-}$& 1.59E-01& -7.29E-06&& -2.77E-03& -7.13E-06&&  4.68E-04& -9.49E-07&&  2.29E-05& -1.87E-07 \\
7p & 1.74E-01& -6.59E-06&& -2.33E-03& -6.80E-06&&  6.17E-04& -8.16E-07&&  5.24E-06& -1.92E-07 \\
$\mathrm{6d_-}$& 6.12E-02& -3.77E-06&& -3.08E-04& -2.20E-05&& -1.41E-03& -1.00E-06&& -9.33E-06& -5.80E-07 \\
6d & 1.08E-01& -3.71E-06&& -4.03E-04& -6.41E-06&& -8.06E-04& -8.36E-07&&  5.40E-07& -4.62E-07 \\
$\mathrm{5f_-}$& 9.73E-02& -1.65E-06&& -2.31E-03& -9.94E-06&&  4.06E-03& -3.88E-07&&  4.95E-05& -1.93E-07 \\
5f & 1.13E-01& -1.73E-06&& -7.25E-04& -5.95E-06&&  2.23E-03& -3.21E-07&&  1.99E-05& -1.76E-07 \\
$\mathrm{5g_-}$& 5.00E-01&  1.66E-06&& -3.60E-02& -4.65E-04&& -3.18E-02& -1.38E-06&& -8.93E-04& -1.88E-06 \\
5g & 5.06E-01&  4.27E-06&& -4.14E-02& -3.23E-05&& -2.05E-02& -8.61E-07&& -5.73E-04& -1.25E-06 \\ \hline
\multicolumn{12}{c}{{\sc self-consistency} in case 99.5\%} \\
7s & 1.93E-02& 6.49E-06&& 1.15E-04& 2.95E-07&& 5.46E-06& 5.36E-07&& 1.50E-07& 1.67E-08  \\
$\mathrm{7p_-}$& 1.26E-02& 6.41E-06&& 1.30E-04& 1.92E-07&& 8.84E-05& 9.95E-07&& 1.45E-06& 8.38E-08  \\
7p & 2.13E-02& 8.14E-06&& 8.27E-05& 2.71E-07&& 1.09E-04& 1.34E-06&& 6.88E-07& 5.94E-08  \\
$\mathrm{6d_-}$& 3.16E-02& 6.06E-06&& 6.97E-05& 3.20E-07&& 5.33E-05& 3.97E-07&& 1.13E-06& 2.56E-08  \\
6d & 4.21E-02& 7.14E-06&& 6.84E-05& 3.70E-07&& 6.56E-05& 4.99E-07&& 4.49E-07& 8.16E-08  \\
$\mathrm{5f_-}$& 2.26E-02& 2.02E-06&& 1.08E-04& 1.12E-07&& 2.66E-04& 2.33E-07&& 1.31E-05& 1.21E-07  \\
5f & 2.79E-02& 2.46E-06&& 2.86E-05& 1.20E-07&& 2.97E-04& 4.18E-07&& 1.96E-06& 2.39E-08  \\
$\mathrm{5g_-}$& 8.96E-03& 1.20E-06&& 3.90E-03& 1.45E-07&& 7.09E-04& 1.16E-06&& 1.02E-05& 7.05E-08  \\
5g & 1.00E-02& 1.55E-06&& 4.84E-03& 1.13E-07&& 7.95E-04& 1.14E-06&& 2.10E-05& 2.03E-07  \\
\multicolumn{12}{c}{{\sc norm-1} in case 99.5\%} \\                                                          
7s & 3.92E-01& -7.81E-05&& -1.28E-03& -3.54E-06&& -5.68E-05&  6.43E-06&&  1.67E-06&  8.67E-08  \\
$\mathrm{7p_-}$& 1.61E-01& -7.22E-05&& -1.43E-03& -2.33E-06&&  6.94E-04&  1.03E-05&&  1.50E-05&  8.08E-07  \\
7p & 1.76E-01& -6.60E-05&& -6.09E-04& -2.17E-06&&  7.30E-04&  1.09E-05&&  5.30E-06&  4.38E-07  \\
$\mathrm{6d_-}$& 8.43E-02& -3.20E-05&& -2.60E-04& -1.95E-06&& -1.71E-04& -2.88E-07&&  3.55E-06& -5.37E-08  \\
6d & 9.89E-02& -3.10E-05&& -1.17E-04& -1.83E-06&& -1.76E-04& -5.19E-07&&  1.40E-06& -2.68E-07  \\
$\mathrm{5f_-}$& 9.88E-02& -1.63E-05&& -7.45E-04& -8.47E-07&&  2.16E-03&  1.95E-06&&  9.44E-05&  9.55E-07  \\
5f & 1.14E-01& -1.71E-05&& -9.50E-05& -7.93E-07&&  2.06E-03&  3.05E-06&&  1.15E-05& -5.87E-08  \\
$\mathrm{5g_-}$& 5.07E-01&  4.29E-05&& -4.09E-02& -4.16E-06&& -7.40E-03& -2.52E-05&& -2.05E-04& -1.42E-06  \\
5g & 5.08E-01&  4.94E-05&& -4.02E-02& -2.81E-06&& -8.22E-03& -2.23E-05&& -4.16E-04& -4.00E-06  \\ \hline
\multicolumn{12}{c}{{\sc self-consistency} in case 99.95\%} \\
7s &  1.94E-02& 1.75E-06&& 2.75E-05& 1.79E-07&& 1.35E-06& 4.58E-08&& 4.49E-08& 1.01E-08 \\
$\mathrm{7p_-}$&  1.28E-02& 1.37E-06&& 2.53E-05& 1.27E-07&& 1.97E-05& 2.34E-08&& 2.43E-07& 7.49E-09 \\
7p &  2.13E-02& 1.76E-06&& 6.87E-05& 1.53E-07&& 1.38E-05& 1.78E-08&& 6.42E-08& 3.92E-09 \\
$\mathrm{6d_-}$&  3.38E-02& 1.23E-06&& 1.79E-04& 2.51E-07&& 8.53E-06& 1.79E-07&& 4.04E-08& 1.48E-09 \\
6d &  4.15E-02& 3.24E-06&& 2.51E-04& 1.85E-07&& 1.17E-05& 1.00E-07&& 1.31E-07& 1.60E-09 \\
$\mathrm{5f_-}$&  2.27E-02& 7.84E-07&& 2.80E-04& 8.41E-08&& 5.83E-06& 1.10E-07&& 8.15E-08& 3.00E-09 \\
5f &  2.80E-02& 1.16E-06&& 3.26E-04& 5.49E-08&& 2.16E-05& 5.17E-09&& 7.15E-07& 4.76E-09 \\
$\mathrm{5g_-}$&  9.15E-03& 1.73E-06&& 5.88E-03& 1.06E-07&& 3.92E-04& 2.19E-07&& 7.32E-07& 6.65E-09 \\
5g &  1.03E-02& 2.95E-06&& 1.56E-03& 2.89E-08&& 4.87E-05& 7.88E-09&& 6.51E-08& 1.33E-09 \\
\multicolumn{12}{c}{{\sc norm-1} in case 99.95\%} \\
7s &  3.93E-01& -2.07E-05&& -2.80E-04& -2.16E-06&& -1.58E-05& -5.74E-07&& -5.60E-07& -7.78E-08 \\
$\mathrm{7p_-}$&  1.61E-01& -1.55E-05&& -2.35E-04& -1.47E-06&&  1.53E-04& -3.07E-07&&  2.19E-06& -1.29E-08 \\
7p &  1.75E-01& -1.41E-05&&  2.65E-04& -1.16E-06&&  1.06E-04& -8.40E-09&&  3.84E-08&  2.13E-09 \\
$\mathrm{6d_-}$&  9.46E-02& -6.52E-06&& -7.54E-04& -1.51E-06&& -5.04E-06& -9.93E-07&& -1.77E-07& -2.39E-08 \\
6d &  9.57E-02& -1.62E-05&& -1.08E-03& -9.25E-07&&  1.19E-05& -4.52E-07&& -3.58E-07&  5.35E-10 \\
$\mathrm{5f_-}$&  9.95E-02& -6.06E-06&&  2.25E-03& -6.12E-07&&  3.98E-05& -7.68E-07&& -4.43E-07& -2.38E-08 \\
5f &  1.14E-01& -7.27E-06&&  2.26E-03& -3.64E-07&&  1.38E-04&  2.05E-08&&  4.57E-06&  3.06E-08 \\
$\mathrm{5g_-}$&  5.13E-01&  7.14E-05&& -4.42E-02& -2.58E-06&& -6.85E-03& -4.94E-06&& -1.51E-05& -1.59E-07 \\
5g &  5.19E-01& -1.01E-04&& -1.92E-02& -6.30E-07&&  9.48E-04&  1.40E-07&&  8.64E-07&  2.64E-08 \\ \hline
\multicolumn{12}{c}{{\sc self-consistency} in case 100\%} \\
7s & 1.94E-02& 4.89E-07&& 1.88E-07& 2.35E-08&& 1.36E-08& 8.83E-09&& 9.51E-09& 9.53E-09 \\
$\mathrm{7p_-}$& 1.28E-02& 4.18E-07&& 1.35E-07& 1.69E-08&& 1.03E-08& 6.14E-09&& 6.13E-09& 6.09E-09 \\
7p & 2.13E-02& 5.04E-07&& 3.60E-07& 1.30E-08&& 7.12E-09& 1.75E-09&& 1.11E-09& 9.85E-10 \\
$\mathrm{6d_-}$& 3.24E-02& 4.18E-07&& 4.74E-07& 5.52E-08&& 2.02E-08& 3.56E-09&& 1.70E-09& 9.10E-10 \\
6d & 4.15E-02& 5.01E-07&& 6.14E-07& 4.81E-08&& 1.83E-08& 2.93E-09&& 1.53E-09& 4.09E-10 \\
$\mathrm{5f_-}$& 2.28E-02& 1.29E-07&& 1.33E-06& 3.12E-08&& 1.01E-08& 1.70E-09&& 7.70E-10& 1.74E-10 \\
5f & 2.80E-02& 1.48E-07&& 1.66E-06& 1.33E-08&& 5.49E-09& 1.00E-09&& 7.17E-10& 1.62E-10 \\
$\mathrm{5g_-}$& 9.34E-03& 7.80E-08&& 1.26E-04& 6.54E-08&& 2.30E-08& 2.62E-09&& 8.96E-10& 2.35E-10 \\
5g & 1.04E-02& 7.64E-08&& 8.66E-05& 2.27E-08&& 8.22E-09& 9.91E-10&& 6.20E-10& 1.70E-10 \\
\multicolumn{12}{c}{{\sc norm-1} in case 100\%} \\
7s & 3.93E-01& -5.84E-06&&  9.35E-07& -3.17E-07&& -1.98E-07& -1.08E-07&& -9.15E-08&-9.11E-08 \\
$\mathrm{7p_-}$& 1.61E-01& -4.62E-06&& -9.80E-07& -2.15E-07&& -1.32E-07& -6.79E-08&& -5.58E-08&-5.52E-08 \\
7p & 1.75E-01& -3.91E-06&& -1.72E-06& -1.20E-07&& -7.58E-08& -3.25E-08&& -2.36E-08&-2.34E-08 \\
$\mathrm{6d_-}$& 8.79E-02& -2.45E-06&& -1.03E-06& -3.35E-07&& -1.36E-07& -3.99E-08&& -2.84E-08&-2.30E-08 \\
6d & 9.59E-02& -2.44E-06&& -1.77E-06& -2.27E-07&& -8.93E-08& -1.66E-08&& -9.43E-09&-3.20E-09 \\
$\mathrm{5f_-}$& 9.98E-02& -9.94E-07&& -2.54E-06& -2.18E-07&& -7.45E-08& -1.60E-08&& -8.72E-09&-3.90E-09 \\
5f & 1.14E-01& -9.86E-07&& -7.48E-06& -7.72E-08&& -3.42E-08& -6.92E-09&& -4.82E-09& 5.80E-10 \\
$\mathrm{5g_-}$& 5.26E-01& -1.88E-06&& -1.58E-03& -1.47E-06&& -5.10E-07& -7.15E-08&& -3.15E-08&-1.50E-08 \\
5g & 5.28E-01& -1.66E-06&& -1.90E-04& -4.56E-07&& -1.64E-07& -2.25E-08&& -1.47E-08& 1.37E-09 \\ \hline
\multicolumn{12}{c}{{\sc self-consistency} in case regular} \\
7s & 1.94E-02& 4.89E-07  \\
$\mathrm{7p_-}$& 1.28E-02& 4.18E-07  \\
7p & 2.13E-02& 5.04E-07  \\
$\mathrm{6d_-}$& 3.24E-02& 4.16E-07  \\
6d & 4.15E-02& 5.00E-07  \\
$\mathrm{5f_-}$& 2.28E-02& 1.29E-07  \\
5f & 2.80E-02& 1.47E-07  \\
$\mathrm{5g_-}$& 9.34E-03& 7.36E-08  \\
5g & 1.04E-02& 7.39E-08  \\
\multicolumn{12}{c}{{\sc norm-1} in case regular} \\ 
7s &  3.93E-01& -5.83E-06 \\
$\mathrm{7p_-}$&  1.61E-01& -4.62E-06 \\
7p &  1.75E-01& -3.90E-06 \\
$\mathrm{6d_-}$&  8.79E-02& -2.44E-06 \\
6d &  9.59E-02& -2.43E-06 \\
$\mathrm{5f_-}$&  9.97E-02& -9.92E-07 \\
5f &  1.14E-01& -9.83E-07 \\
$\mathrm{5g_-}$&  5.26E-01& -1.73E-06 \\
5g &  5.28E-01& -1.57E-06 \\ 
\end{longtable}
}

{\footnotesize
\begin{longtable}{l r r r r r r r r r r r}
\caption{\label{odd_1L_selfconsistency} {\sc self-consistency} and {\sc norm-1} parameters solving the MCDHF equations of the $OS_1$ for odd parity states in cases 95, 100\% and regular.
Columns with 'x' var. means the number of variations with constructed CSF basis. 
In. and Fin. means the initial and final results of these parameters solving the MCDHF equations.}\\
\hline\hline
\multicolumn{1}{c}{Subshell}& \multicolumn{2}{c}{1 var.} && \multicolumn{2}{c}{2 var.} && \multicolumn{2}{c}{3 var.} && \multicolumn{2}{c}{4 var.}  \\
\cline{2-3} \cline{5-6} \cline{8-9} \cline{11-12}
& \multicolumn{1}{c}{In.} & \multicolumn{1}{c}{Fin.} && \multicolumn{1}{c}{In.} & \multicolumn{1}{c}{Fin.} && \multicolumn{1}{c}{In.} & \multicolumn{1}{c}{Fin.} && \multicolumn{1}{c}{In.} & \multicolumn{1}{c}{Fin.}   \\
\hline
\noalign{\smallskip}
\endfirsthead
\caption{Continued.}\\
\hline\hline
\multicolumn{1}{c}{Subshell}& \multicolumn{2}{c}{1 var.} && \multicolumn{2}{c}{2 var.} && \multicolumn{2}{c}{3 var.} && \multicolumn{2}{c}{4 var.}  \\
\cline{2-3} \cline{5-6} \cline{8-9} \cline{11-12}
& \multicolumn{1}{c}{In.} & \multicolumn{1}{c}{Fin.} && \multicolumn{1}{c}{In.} & \multicolumn{1}{c}{Fin.} && \multicolumn{1}{c}{In.} & \multicolumn{1}{c}{Fin.} && \multicolumn{1}{c}{In.} & \multicolumn{1}{c}{Fin.}   \\
\hline
\noalign{\smallskip}
\endhead
\noalign{\smallskip}
\hline
\hline
\endfoot
\noalign{\smallskip}
\multicolumn{12}{c}{{\sc self-consistency} in case 95\%} \\
7s & 1.92E-02& 1.08E-07&& 4.18E-04& 4.77E-07&& 8.10E-06& 1.39E-06&& 8.11E-07& 5.80E-08 \\
$\mathrm{7p_-}$& 1.16E-02& 8.61E-07&& 1.26E-03& 2.79E-07&& 9.60E-05& 7.68E-07&& 1.23E-05& 1.55E-08 \\
7p & 1.78E-02& 3.04E-07&& 9.97E-04& 4.09E-07&& 6.92E-05& 1.07E-06&& 7.17E-06& 5.51E-08 \\
$\mathrm{6d_-}$& 1.83E-01& 2.68E-07&& 8.83E-04& 4.75E-07&& 2.89E-04& 2.99E-06&& 3.19E-05& 2.29E-07 \\
6d & 1.94E-01& 3.63E-07&& 8.31E-04& 1.92E-06&& 1.81E-04& 2.26E-06&& 3.81E-05& 1.37E-07 \\
$\mathrm{5f_-}$& 1.99E-02& 2.38E-07&& 1.05E-03& 7.91E-08&& 2.60E-04& 1.17E-06&& 9.20E-05& 9.07E-08 \\
5f & 2.27E-02& 2.36E-07&& 8.93E-04& 7.19E-07&& 3.06E-04& 4.88E-07&& 1.30E-04& 6.20E-08 \\
$\mathrm{5g_-}$& 7.07E-03& 2.94E-09&& 2.59E-03& 4.39E-06&& 4.98E-03& 4.11E-06&& 3.18E-04& 3.76E-07 \\
5g & 7.90E-03& 2.06E-09&& 5.45E-03& 4.60E-06&& 2.14E-03& 7.15E-07&& 5.45E-05& 5.56E-08 \\
\multicolumn{12}{c}{{\sc norm-1} in case 95\%} \\                                                                                                             
7s & 3.65E-01& -1.45E-06&& -4.76E-03& -5.55E-06&& -7.63E-05& -1.60E-05&&  9.20E-06& -7.06E-07 \\
$\mathrm{7p_-}$& 1.79E-01& -5.42E-06&& -8.38E-03& -3.17E-06&& -8.29E-04& -8.94E-06&&  7.31E-05&  4.21E-08 \\
7p & 1.89E-01& -2.05E-06&& -6.57E-03& -3.26E-06&& -5.06E-04& -8.06E-06&&  2.95E-05&  3.04E-07 \\
$\mathrm{6d_-}$& 1.83E+00& -9.93E-07&& -4.95E-03& -3.05E-06&& -1.37E-03& -1.78E-05&& -1.47E-04& -1.26E-06 \\
6d & 1.46E+00& -1.42E-06&& -2.16E-03& -9.17E-06&& -8.14E-04& -1.10E-05&& -1.35E-04& -5.60E-07 \\
$\mathrm{5f_-}$& 4.72E-02& -1.85E-06&& -6.49E-03& -3.34E-07&&  1.89E-03& -7.85E-06&&  6.27E-04& -5.76E-07 \\
5f & 3.05E-02& -1.54E-06&& -4.18E-03& -3.97E-06&&  1.97E-03& -2.94E-06&&  8.32E-04&  4.01E-07 \\
$\mathrm{5g_-}$& 8.36E-01&  9.30E-08&& -3.50E-02& -1.34E-04&& -2.75E-02& -1.08E-04&& -4.02E-03& -9.23E-06 \\
5g & 8.22E-01&  5.39E-08&& -1.03E-02& -1.27E-04&& -3.20E-02& -1.53E-05&& -1.21E-03& -1.13E-06 \\ \hline
\multicolumn{12}{c}{{\sc self-consistency} in case 100\%} \\
7s & 2.06E-02& 1.46E-07&& 3.39E-07& 5.99E-08&& 3.10E-08& 9.52E-09&& 7.99E-09& 8.46E-09 \\
$\mathrm{7p_-}$& 1.31E-02& 1.21E-07&& 2.23E-07& 3.39E-08&& 1.90E-08& 6.53E-09&& 5.31E-09& 5.26E-09 \\
7p & 1.97E-02& 1.36E-07&& 7.69E-07& 3.39E-08&& 1.83E-08& 4.13E-09&& 1.90E-09& 1.01E-09 \\
$\mathrm{6d_-}$& 1.47E-01& 1.37E-07&& 4.23E-07& 1.59E-07&& 6.22E-08& 9.39E-09&& 3.69E-09& 1.69E-09 \\
6d & 1.68E-01& 1.73E-07&& 5.19E-07& 1.43E-07&& 5.62E-08& 8.55E-09&& 3.20E-09& 1.22E-09 \\
$\mathrm{5f_-}$& 2.16E-02& 4.63E-08&& 1.35E-06& 7.82E-08&& 2.80E-08& 3.64E-09&& 1.35E-09& 5.35E-10 \\
5f & 2.40E-02& 5.77E-08&& 1.61E-06& 3.46E-08&& 1.42E-08& 2.18E-09&& 8.58E-10& 3.28E-10 \\
$\mathrm{5g_-}$& 8.94E-03& 4.19E-08&& 1.35E-04& 2.45E-07&& 7.65E-08& 7.04E-09&& 2.24E-09& 7.69E-10 \\
5g & 9.77E-03& 4.81E-08&& 8.24E-05& 1.01E-07&& 2.93E-08& 2.86E-09&& 9.56E-10& 3.26E-10 \\
\multicolumn{12}{c}{{\sc norm-1} in case 100\%} \\                                                    
7s & 3.77E-01& -1.69E-06&&  2.44E-06& -7.16E-07&& -3.95E-07& -1.43E-07&& -1.08E-07& -9.36E-08  \\
$\mathrm{7p_-}$& 1.98E-01& -1.24E-06&& -1.29E-06& -4.06E-07&& -2.32E-07& -8.90E-08&& -6.73E-08& -5.80E-08  \\
7p & 2.00E-01& -1.08E-06&& -3.43E-06& -2.81E-07&& -1.62E-07& -5.15E-08&& -3.39E-08& -2.63E-08  \\
$\mathrm{6d_-}$& 1.28E+00& -8.67E-07&& -8.32E-07& -9.26E-07&& -3.78E-07& -7.50E-08&& -4.16E-08& -2.94E-08  \\
6d & 1.15E+00& -8.61E-07&& -1.11E-06& -6.61E-07&& -2.66E-07& -4.35E-08&& -1.80E-08& -8.51E-09  \\
$\mathrm{5f_-}$& 5.71E-02& -3.29E-07&& -3.17E-06& -4.88E-07&& -1.82E-07& -2.68E-08&& -1.18E-08& -6.40E-09  \\
5f & 3.72E-02& -3.51E-07&& -6.58E-06& -1.78E-07&& -7.81E-08& -1.29E-08&& -5.28E-09& -2.14E-09  \\
$\mathrm{5g_-}$& 1.05E+00& -9.23E-07&& -1.84E-03& -5.08E-06&& -1.60E-06& -1.56E-07&& -5.66E-08& -2.57E-08  \\
5g & 1.03E+00& -9.39E-07&& -4.28E-04& -1.85E-06&& -5.42E-07& -5.54E-08&& -1.97E-08& -7.67E-09  \\ \hline
\multicolumn{12}{c}{{\sc self-consistency} in case regular} \\                                               
7s & 2.06E-02& 1.45E-07  \\
$\mathrm{7p_-}$& 1.31E-02& 1.21E-07  \\
7p & 1.97E-02& 1.36E-07  \\
$\mathrm{6d_-}$& 1.47E-01& 1.36E-07  \\
6d & 1.68E-01& 1.71E-07  \\
$\mathrm{5f_-}$& 2.16E-02& 4.57E-08  \\
5f & 2.40E-02& 5.75E-08  \\
$\mathrm{5g_-}$& 8.94E-03& 3.86E-08  \\
5g & 9.77E-03& 4.67E-08  \\
\multicolumn{12}{c}{{\sc norm-1} in case regular} \\ 
7s & 3.77E-01& -1.68E-06 \\
$\mathrm{7p_-}$& 1.98E-01& -1.24E-06 \\
7p & 2.00E-01& -1.08E-06 \\
$\mathrm{6d_-}$& 1.28E+00& -8.57E-07 \\
6d & 1.15E+00& -8.54E-07 \\
$\mathrm{5f_-}$& 5.71E-02& -3.25E-07 \\
5f & 3.72E-02& -3.50E-07 \\
$\mathrm{5g_-}$& 1.05E+00& -8.35E-07 \\
5g & 1.03E+00& -8.94E-07 \\ 
\end{longtable}
}

{\footnotesize
\begin{longtable}{l r r r r r r r r r r r}
\caption{\label{even_2L_selfconsistency} {\sc self-consistency} and {\sc norm-1} parameters solving the MCDHF equations of the $OS_2$ for even parity states in cases 95, 99, 99.5, 99.95\% and regular.
Columns with 'x' var. means the number of variations with constructed CSF basis. 
In. and Fin. means the initial and final results of these parameters solving the MCDHF equations.}\\
\hline\hline
\multicolumn{1}{c}{Subshell}& \multicolumn{2}{c}{1 var.} && \multicolumn{2}{c}{2 var.} && \multicolumn{2}{c}{3 var.} && \multicolumn{2}{c}{4 var.}  \\
\cline{2-3} \cline{5-6} \cline{8-9} \cline{11-12}
& \multicolumn{1}{c}{In.} & \multicolumn{1}{c}{Fin.} && \multicolumn{1}{c}{In.} & \multicolumn{1}{c}{Fin.} && \multicolumn{1}{c}{In.} & \multicolumn{1}{c}{Fin.} && \multicolumn{1}{c}{In.} & \multicolumn{1}{c}{Fin.}   \\
\hline
\noalign{\smallskip}
\endfirsthead
\caption{Continued.}\\
\hline\hline
\multicolumn{1}{c}{Subshell}& \multicolumn{2}{c}{1 var.} && \multicolumn{2}{c}{2 var.} && \multicolumn{2}{c}{3 var.} && \multicolumn{2}{c}{4 var.}  \\
\cline{2-3} \cline{5-6} \cline{8-9} \cline{11-12}
& \multicolumn{1}{c}{In.} & \multicolumn{1}{c}{Fin.} && \multicolumn{1}{c}{In.} & \multicolumn{1}{c}{Fin.} && \multicolumn{1}{c}{In.} & \multicolumn{1}{c}{Fin.} && \multicolumn{1}{c}{In.} & \multicolumn{1}{c}{Fin.}   \\
\hline
\noalign{\smallskip}
\endhead
\noalign{\smallskip}
\hline
\hline
\endfoot
\noalign{\smallskip}
\multicolumn{12}{c}{{\sc self-consistency} in case 95\%} \\
8s & 1.30E-02& 2.78E-06&& 1.44E-03& 1.13E-06&& 1.10E-04& 4.00E-07&& 1.98E-06& 2.99E-07 \\
$\mathrm{8p_-}$& 2.00E-02& 3.11E-06&& 4.88E-03& 1.17E-06&& 1.37E-04& 3.17E-07&& 3.15E-06& 4.07E-07 \\
8p & 2.34E-02& 3.33E-06&& 4.32E-03& 1.52E-06&& 1.81E-04& 4.38E-07&& 8.81E-06& 3.75E-07 \\
$\mathrm{7d_-}$& 1.62E-02& 4.43E-06&& 1.92E-03& 2.12E-06&& 2.47E-04& 1.38E-06&& 8.28E-05& 1.66E-06 \\
7d & 2.00E-02& 6.67E-06&& 2.54E-03& 2.79E-06&& 2.10E-04& 1.19E-06&& 1.11E-05& 6.28E-07 \\
$\mathrm{6f_-}$& 2.48E-02& 7.65E-07&& 1.64E-03& 1.54E-06&& 6.15E-04& 6.54E-06&& 2.21E-04& 9.37E-06 \\
6f & 2.89E-02& 1.99E-06&& 1.45E-03& 3.01E-06&& 6.99E-04& 1.09E-05&& 1.24E-04& 6.61E-06 \\
$\mathrm{6g_-}$& 2.53E-03& 1.29E-08&& 7.80E-03& 2.10E-08&& 8.40E-04& 2.05E-07&& 1.14E-04& 1.88E-07 \\
6g & 2.90E-03& 2.51E-08&& 9.87E-03& 5.07E-08&& 1.14E-03& 3.92E-07&& 6.57E-05& 4.17E-07 \\
\multicolumn{12}{c}{{\sc norm-1} in case 95\%} \\
8s & 2.77E+00& -1.82E-04&& -6.46E-02& -7.35E-05&& -7.06E-03& -2.54E-05&& -1.11E-04& -1.88E-05 \\
$\mathrm{8p_-}$& 4.62E+00& -2.33E-04&& -9.37E-03& -7.25E-05&& -7.56E-03&  5.53E-06&& -1.91E-04&  1.44E-05 \\
8p & 3.89E+00& -1.59E-04&& -4.01E-02& -5.88E-05&& -6.25E-03&  2.44E-06&& -2.68E-04&  6.65E-06 \\
$\mathrm{7d_-}$& 6.39E+00& -9.60E-05&& -3.47E-02& -4.91E-05&& -6.43E-03& -4.04E-05&& -2.25E-03& -4.96E-05 \\
7d & 6.49E+00& -1.14E-04&& -3.47E-02& -5.31E-05&& -1.96E-03& -2.61E-05&&  5.77E-07& -1.16E-05 \\
$\mathrm{6f_-}$& 2.26E+00& -1.77E-05&& -2.90E-02& -3.32E-05&& -1.38E-02& -1.50E-04&& -5.39E-03& -2.17E-04 \\
6f & 2.24E+00& -3.83E-05&& -2.40E-02& -5.83E-05&& -1.41E-02& -2.23E-04&& -2.47E-03& -1.36E-04 \\
$\mathrm{6g_-}$& 2.32E+00& -6.24E-07&&  1.93E-01& -4.76E-07&& -1.29E-02& -5.42E-06&& -2.73E-03& -4.43E-06 \\
6g & 2.34E+00& -1.90E-06&&  2.22E-01& -1.06E-06&& -1.76E-02& -9.07E-06&& -1.56E-03& -9.76E-06 \\ \hline
\multicolumn{12}{c}{{\sc self-consistency} in case 99.95\%} \\
8s & 1.27E-02& 2.00E-06&& 6.35E-04& 1.12E-07&& 7.33E-06& 1.78E-07&& 2.02E-07& 8.76E-08 \\
$\mathrm{8p_-}$& 4.49E-02& 1.22E-06&& 1.37E-04& 1.51E-07&& 7.95E-06& 2.51E-07&& 2.77E-07& 1.19E-07 \\
8p & 4.23E-02& 1.64E-06&& 1.68E-04& 2.57E-07&& 5.89E-06& 4.78E-07&& 4.48E-07& 2.95E-07 \\
$\mathrm{7d_-}$& 1.78E-02& 2.34E-06&& 2.53E-04& 6.28E-07&& 4.24E-06& 9.84E-07&& 8.12E-07& 6.07E-07 \\
7d & 2.18E-02& 3.05E-06&& 2.10E-04& 7.80E-07&& 5.18E-06& 1.26E-06&& 9.68E-07& 9.19E-07 \\
$\mathrm{6f_-}$& 2.63E-02& 1.21E-06&& 8.76E-04& 3.36E-06&& 1.22E-05& 4.67E-06&& 3.00E-06& 2.41E-06 \\
6f & 3.01E-02& 1.52E-06&& 9.79E-04& 3.93E-06&& 1.33E-05& 5.62E-06&& 5.03E-06& 3.84E-06 \\
$\mathrm{6g_-}$& 8.88E-03& 2.13E-07&& 1.49E-03& 1.07E-05&& 1.40E-04& 1.51E-05&& 2.23E-05& 6.71E-06 \\
6g & 1.00E-02& 1.33E-07&& 1.56E-03& 1.27E-05&& 1.76E-04& 1.93E-05&& 5.99E-05& 1.69E-05 \\
\multicolumn{12}{c}{{\sc norm-1} in case 99.95\%} \\
8s & 2.71E+00& -8.54E-05&& -3.99E-03&  2.44E-06&& -2.48E-04&  4.17E-06&&  3.15E-06&  1.69E-06 \\
$\mathrm{8p_-}$& 1.22E+01& -6.57E-05&& -3.70E-03&  5.97E-06&&  3.09E-04&  1.01E-05&&  1.31E-05&  4.09E-06 \\
8p & 6.02E+00& -5.35E-05&& -5.57E-03&  6.31E-06&&  1.78E-04&  1.38E-05&&  1.61E-05&  8.44E-06 \\
$\mathrm{7d_-}$& 7.52E+00& -5.59E-05&& -6.65E-03& -1.27E-05&&  3.82E-05& -2.05E-05&& -1.64E-05& -1.24E-05 \\
7d & 7.70E+00& -5.93E-05&& -1.83E-03& -1.24E-05&&  4.84E-05& -2.10E-05&& -1.51E-05& -1.58E-05 \\
$\mathrm{6f_-}$& 2.29E+00& -2.60E-05&& -1.81E-02& -7.60E-05&&  2.51E-06& -1.07E-04&& -7.22E-05& -5.54E-05 \\
6f & 2.26E+00& -2.86E-05&& -1.78E-02& -7.78E-05&& -1.59E-04& -1.12E-04&& -1.03E-04& -7.69E-05 \\
$\mathrm{6g_-}$& 1.41E+00&  1.02E-05&& -5.35E-02& -8.23E-04&& -1.10E-02& -1.22E-03&& -1.30E-03& -5.44E-04 \\
6g & 1.41E+00&  5.52E-06&& -5.17E-02& -8.79E-04&& -1.25E-02& -1.41E-03&& -3.49E-03& -1.26E-03 \\ \hline
\multicolumn{12}{c}{{\sc self-consistency} in case regular} \\
8s & 1.28E-02& 1.81E-07  \\
$\mathrm{8p_-}$& 2.78E-02& 2.98E-07  \\
8p & 5.61E-02& 4.20E-07  \\
$\mathrm{7d_-}$& 1.83E-02& 9.36E-07  \\
7d & 2.22E-02& 1.14E-06  \\
$\mathrm{6f_-}$& 2.67E-02& 4.32E-06  \\
6f & 3.06E-02& 4.99E-06  \\
$\mathrm{6g_-}$& 9.08E-03& 1.41E-05  \\
6g & 1.01E-02& 1.68E-05  \\
\multicolumn{12}{c}{{\sc norm-1} in case regular} \\ 
8s & 2.72E+00&  2.38E-06 \\
$\mathrm{8p_-}$& 6.76E+00&  1.29E-05 \\
8p & 8.44E+00&  1.17E-05 \\
$\mathrm{7d_-}$& 7.66E+00& -2.02E-05 \\
7d & 7.81E+00& -1.93E-05 \\
$\mathrm{6f_-}$& 2.31E+00& -9.90E-05 \\
6f & 2.28E+00& -9.98E-05 \\
$\mathrm{6g_-}$& 1.42E+00& -1.15E-03 \\
6g & 1.42E+00& -1.23E-03 \\ 
\end{longtable}
}

{\footnotesize
\begin{longtable}{l r r r r r r r r r r r}
\caption{\label{odd_2L_selfconsistency} {\sc self-consistency} and {\sc norm-1} parameters solving the MCDHF equations of the $OS_2$ for odd parity states in cases 95, 99.95\% and regular.
Columns with 'x' var. means the number of variations with constructed CSF basis. 
In. and Fin. means the initial and final results of these parameters solving the MCDHF equations.}\\
\hline\hline
\multicolumn{1}{c}{Subshell}& \multicolumn{2}{c}{1 var.} && \multicolumn{2}{c}{2 var.} && \multicolumn{2}{c}{3 var.} && \multicolumn{2}{c}{4 var.}  \\
\cline{2-3} \cline{5-6} \cline{8-9} \cline{11-12}
& \multicolumn{1}{c}{In.} & \multicolumn{1}{c}{Fin.} && \multicolumn{1}{c}{In.} & \multicolumn{1}{c}{Fin.} && \multicolumn{1}{c}{In.} & \multicolumn{1}{c}{Fin.} && \multicolumn{1}{c}{In.} & \multicolumn{1}{c}{Fin.}   \\
\hline
\noalign{\smallskip}
\endfirsthead
\caption{Continued.}\\
\hline\hline
\multicolumn{1}{c}{Subshell}& \multicolumn{2}{c}{1 var.} && \multicolumn{2}{c}{2 var.} && \multicolumn{2}{c}{3 var.} && \multicolumn{2}{c}{4 var.}  \\
\cline{2-3} \cline{5-6} \cline{8-9} \cline{11-12}
& \multicolumn{1}{c}{In.} & \multicolumn{1}{c}{Fin.} && \multicolumn{1}{c}{In.} & \multicolumn{1}{c}{Fin.} && \multicolumn{1}{c}{In.} & \multicolumn{1}{c}{Fin.} && \multicolumn{1}{c}{In.} & \multicolumn{1}{c}{Fin.}   \\
\hline
\noalign{\smallskip}
\endhead
\noalign{\smallskip}
\hline
\hline
\endfoot
\noalign{\smallskip}
\multicolumn{12}{c}{{\sc self-consistency} in case 95\%} \\
8s & 1.92E-02& 4.01E-06&& 1.98E-03& 1.43E-06&& 1.65E-04& 7.28E-07&& 4.93E-06& 1.19E-06 \\
$\mathrm{8p_-}$& 1.09E-02& 2.59E-06&& 1.74E-03& 2.10E-06&& 2.12E-04& 9.39E-07&& 1.54E-05& 1.40E-06 \\
8p & 1.63E-02& 3.55E-06&& 4.05E-03& 3.37E-06&& 1.63E-02& 1.58E-06&& 1.74E-05& 2.05E-06 \\
$\mathrm{7d_-}$& 1.86E-02& 6.64E-06&& 2.29E-03& 3.25E-06&& 5.33E-03& 1.77E-06&& 1.52E-04& 5.03E-06 \\
7d & 2.31E-02& 9.54E-06&& 2.99E-03& 4.47E-06&& 7.52E-03& 2.28E-06&& 4.07E-05& 2.44E-06 \\
$\mathrm{6f_-}$& 2.89E-02& 2.02E-07&& 2.12E-03& 4.43E-07&& 1.11E-04& 3.54E-07&& 1.83E-05& 7.14E-07 \\
6f & 3.34E-02& 8.92E-07&& 2.05E-03& 1.36E-06&& 1.13E-04& 5.80E-07&& 3.85E-05& 1.74E-06 \\
$\mathrm{6g_-}$& 8.53E-03& 6.13E-08&& 1.75E-03& 3.53E-07&& 4.19E-04& 6.46E-07&& 5.46E-05& 2.85E-06 \\
6g & 9.63E-03& 8.86E-08&& 2.16E-03& 8.10E-07&& 3.84E-04& 7.26E-07&& 6.01E-05& 2.01E-06 \\
\multicolumn{12}{c}{{\sc norm-1} in case 95\%} \\
8s & 3.63E+00& -2.58E-04&& -6.68E-02& -8.99E-05&& -9.85E-03& -4.62E-05&& -3.20E-04& -7.49E-05 \\
$\mathrm{8p_-}$& 5.19E+00& -1.85E-04&& -5.97E-02& -1.29E-04&& -1.11E-02& -5.63E-05&& -8.60E-04& -8.37E-05 \\
8p & 5.17E+00& -1.64E-04&& -5.06E-02& -1.22E-04&&  2.82E-01& -5.97E-05&& -5.73E-04& -7.13E-05 \\
$\mathrm{7d_-}$& 5.04E+00& -1.70E-04&& -3.48E-02& -5.82E-05&&  5.12E-02& -3.62E-05&& -2.90E-03& -1.24E-04 \\
7d & 5.17E+00& -1.76E-04&& -3.46E-02& -6.88E-05&&  6.55E-02& -3.72E-05&& -6.31E-04& -3.74E-05 \\
$\mathrm{6f_-}$& 2.78E+00& -3.47E-06&& -3.24E-02& -3.85E-06&&  6.74E-04& -4.65E-06&& -2.17E-04& -1.03E-05 \\
6f & 2.78E+00& -1.31E-05&& -2.87E-02& -1.91E-05&& -1.78E-03& -7.56E-06&& -5.30E-04& -2.62E-05 \\
$\mathrm{6g_-}$& 1.97E+00& -3.76E-06&& -6.62E-02& -1.78E-05&& -2.31E-02& -3.69E-05&& -3.58E-03& -1.70E-04 \\
6g & 2.02E+00& -4.56E-06&& -6.46E-02& -3.66E-05&& -2.00E-02& -3.55E-05&& -2.80E-03& -9.97E-05 \\ \hline
\multicolumn{12}{c}{{\sc self-consistency} in case 99.95\%} \\
8s & 1.53E-02& 3.30E-06&& 2.55E-04& 1.88E-07&& 1.35E-06& 5.54E-07&& 3.44E-07& 5.66E-08 \\
$\mathrm{8p_-}$& 1.20E-02& 2.25E-06&& 2.44E-04& 2.85E-07&& 8.84E-06& 8.65E-07&& 4.29E-07& 5.76E-08 \\
8p & 1.71E-02& 3.40E-06&& 2.40E-04& 4.50E-07&& 4.45E-06& 7.99E-07&& 6.02E-07& 8.28E-08 \\
$\mathrm{7d_-}$& 1.92E-02& 3.28E-06&& 2.99E-04& 1.19E-06&& 2.30E-05& 1.13E-06&& 6.49E-07& 7.88E-08 \\
7d & 2.35E-02& 4.30E-06&& 3.41E-04& 1.09E-06&& 1.65E-05& 1.20E-06&& 7.19E-07& 1.12E-07 \\
$\mathrm{6f_-}$& 3.02E-02& 3.54E-07&& 8.60E-04& 8.35E-06&& 5.17E-06& 1.75E-06&& 1.15E-06& 2.96E-07 \\
6f & 3.46E-02& 4.51E-07&& 9.02E-04& 8.88E-06&& 6.05E-06& 1.27E-06&& 7.21E-07& 1.47E-07 \\
$\mathrm{6g_-}$& 8.90E-03& 1.26E-07&& 1.78E-03& 9.31E-06&& 4.81E-05& 8.40E-06&& 5.37E-06& 9.23E-07 \\
6g & 1.01E-02& 1.71E-07&& 1.79E-03& 1.12E-05&& 1.04E-05& 7.38E-07&& 2.53E-07& 6.98E-08 \\
\multicolumn{12}{c}{{\sc norm-1} in case 99.95\%} \\
8s & 2.66E+00& -1.18E-04&& -6.42E-03& -8.82E-06&& -5.44E-05&  2.21E-05&&  1.43E-05&  2.12E-06 \\
$\mathrm{8p_-}$& 5.34E+00& -1.11E-04&& -9.91E-03&  3.64E-06&&  4.06E-04&  4.27E-05&&  2.14E-05&  2.58E-06 \\
8p & 5.26E+00& -9.00E-05&& -7.23E-03&  6.23E-06&&  1.37E-04&  2.18E-05&&  1.89E-05&  2.33E-06 \\
$\mathrm{7d_-}$& 5.42E+00& -6.43E-05&& -6.71E-03& -2.60E-05&&  3.43E-04&  1.15E-05&&  4.94E-06& -9.51E-08 \\
7d & 5.48E+00& -6.98E-05&& -1.51E-03& -1.64E-05&&  1.99E-04&  1.11E-05&&  5.70E-06&  7.73E-07 \\
$\mathrm{6f_-}$& 2.80E+00& -5.35E-07&& -1.55E-02& -1.59E-04&& -8.71E-05& -3.41E-05&& -2.26E-05& -5.72E-06 \\
6f & 2.81E+00& -1.19E-06&& -1.43E-02& -1.47E-04&& -1.00E-04& -2.19E-05&& -1.24E-05& -2.51E-06 \\
$\mathrm{6g_-}$& 2.03E+00& -6.29E-06&& -6.56E-02& -5.26E-04&& -2.78E-03& -4.85E-04&& -3.20E-04& -5.34E-05 \\
6g & 2.07E+00& -7.34E-06&& -6.15E-02& -5.68E-04&& -1.41E-04& -3.83E-05&& -1.41E-05& -3.59E-06 \\ \hline
\multicolumn{12}{c}{{\sc self-consistency} in case regular} \\
8s & 1.52E-02& 7.14E-05  \\
$\mathrm{8p_-}$& 1.20E-02& 2.44E-06  \\
8p & 1.71E-02& 3.49E-06  \\
$\mathrm{7d_-}$& 1.91E-02& 3.88E-06  \\
7d & 2.34E-02& 4.78E-06  \\
$\mathrm{6f_-}$& 3.05E-02& 5.63E-06  \\
6f & 3.48E-02& 6.17E-06  \\
$\mathrm{6g_-}$& 9.44E-03& 6.03E-06  \\
6g & 1.05E-02& 6.39E-06  \\
\multicolumn{12}{c}{{\sc norm-1} in case regular} \\ 
8s & 2.65E+00& -1.51E-04 \\
$\mathrm{8p_-}$& 5.37E+00& -1.20E-04 \\
8p & 5.28E+00& -1.12E-04 \\
$\mathrm{7d_-}$& 5.50E+00& -8.91E-05 \\
7d & 5.55E+00& -8.59E-05 \\
$\mathrm{6f_-}$& 2.80E+00& -1.05E-04 \\
6f & 2.81E+00& -9.94E-05 \\
$\mathrm{6g_-}$& 2.05E+00& -3.45E-04 \\
6g & 2.12E+00& -3.23E-04 \\
\end{longtable}
} 

Table \ref{CSF_summary} summarizes CSF bases used in the MCDHF computations of $OS_1$ and $OS_2$.
The number of CSFs is given for each variation step using the RSMBPT method with a different fraction, as well as in the regular {\sc Grasp}2018 calculations. It is seen, that the largest change in the number of CSFs occurs after the first variation, when the selection procedure 
is based on radial wavefunctions estimated using the Thomas-Fermi potential.
After a few variations, the CSFs basis almost does not change. 

\begin{table*}[!ht]
\setlength{\tabcolsep}{3pt}
\caption{Number of CSF in the MCDHF computations of $OS_1$ and $OS_2$ using regular way and the RSMBPT method.}            
\label{CSF_summary} 
\centering
\begin{tabular}{l r r r r r r r r r}
\hline\hline
Case &\multicolumn{4}{c}{$OS_1$} && \multicolumn{4}{c}{$OS_2$}  \\
\cline{2-5} \cline{7-10}
 &\multicolumn{1}{c}{1 var.} & \multicolumn{1}{c}{2 var.} & \multicolumn{1}{c}{3 var.} & \multicolumn{1}{c}{4 var.} &&
\multicolumn{1}{c}{1 var.} & \multicolumn{1}{c}{2 var.} & \multicolumn{1}{c}{3 var.} & \multicolumn{1}{c}{4 var.} \\
\hline
\noalign{\smallskip}
\multicolumn{10}{c}{Even} \\
95\% &    508364&  569574&  570467&  571019 &&  733794& 1023336& 1141939& 1147646 \\
99\% &    689625&  785778&  898767&  897052 && 1194590& 1746407& 1806590& 1847202 \\
99.5\% &  772411&  880322&  980963&  981664 && 1398101& 1979519& 2071879& 2111506 \\ 
99.95\% & 966043& 1144631& 1173196& 1173442 && 1944852& 2525688& 2567785& 2567315 \\
100\% &  1287673& 1303671& 1303659& 1303659 &&  \\
Regular& 1303709 & & &                      && 2923523 \\
\hline
\multicolumn{10}{c}{Odd} \\
95\% &   197074& 226517& 245285& 248960 &&  280434& 422802& 439826& 442597 \\
99\% &   269522& 322049& 340528& 340543 &&  465186& 685001& 700179& 700938 \\ 
99.5\% & 303520& 353355& 371760& 371834 &&  550123& 783504& 799867& 799302 \\  
99.95\% &380829& 444754& 450637& 450656 &&  764152& 973620& 974618& 974715 \\ 
100\% &  500591& 507233& 507231& 507231 &&   \\ 
Regular& 507234 & & &                   && 1126622 \\
\hline
\hline
\end{tabular}
\end{table*}

Table \ref{TE_diff_even1L_95} presents the total energies 
from regular \textbf{CV+C+VV~MCDHF} and differences 
between the \textbf{CV+C+VV~MCDHF~(RSMBPT~95\%)} and \textbf{CV+C+VV~MCDHF} calculations. 
In the last lines of the Table the smallest ($\Delta E_{min}$) and the largest ($\Delta E_{max}$)
differences with the results of the regular 
{\sc Grasp}2018 calculations (\textbf{CV+C+VV~MCDHF}), as well as the root-mean-square (rms) are given for each computation. 
This Table shows results for all computed even states of $OS_1$.
As seen from the Table, the results converge and almost do not change after few variations.
The differences between both (regular and RSMBPT~95\%) computations are similar for most levels at the same variation.
The rms in the 'case 95\%' is equal 0.00847151 a.u. at final variation step, 
the $\Delta E_{min}$ is 0.00586482 a.u. (or 0.00024168\%), 
$\Delta E_{max}$ is 0.01143258 a.u. (or 0.00047095\%).
Since the trends are similar, only a summary of the $\Delta E_{min}$, $\Delta E_{max}$ and rms 
is presented in Table \ref{min_max_rms_summary} for the remaining computed cases. 
It is seen, that by including the most significant CV+C+VV correlations -- increasing the amount 
of these correlations (95, 99, 99.5, 99.95 and 100\%) -- the results converge to the regular {\sc Grasp}2018 results, 
and in the case of ‘100\%’, reproduce them. 
Furthermore, the results with the specified amount (95, 99, 99.5, 99.95\%) of correlations also shows good agreement with
the regular {\sc Grasp}2018 data, but the CSFs bases are significantly smaller compared to the regular ones,
which is relevant for complex systems, when CSF bases grow rapidly with each new OS and each opened core shell.
It should be noted, that in some cases, especially when the specified fraction of correlations is smaller,
the order of levels that are very close to each other can differ from the regular computations.

{\footnotesize
\begin{longtable}{r r r r r r r r}
\caption{\label{TE_diff_even1L_95} The total energies (in a.u.) from \textbf{CV+C+VV~MCDHF} calculations and differences (in a.u.) 
between \textbf{CV+C+VV~MCDHF~(RSMBPT~95\%)} and \textbf{CV+C+VV~MCDHF} 
energies ($\Delta E_{\textbf{(CV+C+VV~MCDHF~(RSMBPT~95\%))-(CV+C+VV~MCDHF)}}$) 
for $OS_1$ even states of Se~III are given when CV, C and VV correlations are included in the computations.
'Pos' means numbering of the levels for specific parity and J value in the spectra.}\\
\hline\hline
\multicolumn{1}{c}{\multirow{2}{*}{No}} & \multicolumn{1}{c}{\multirow{2}{*}{Pos}} & \multicolumn{1}{c}{\multirow{2}{*}{$J$}} & \multicolumn{1}{c}{\multirow{2}{*}{\textbf{CV+C+VV~MCDHF}}} & \multicolumn{4}{c}{$\Delta E_{\textbf{(CV+C+VV~MCDHF~(RSMBPT~95\%))-(CV+C+VV~MCDHF)}}$}\\
\cline{5-8}
&&&&\multicolumn{1}{c}{1 var.} & \multicolumn{1}{c}{2 var.} & \multicolumn{1}{c}{3 var.} & \multicolumn{1}{c}{4 var.}  \\
\hline
\noalign{\smallskip}
\endfirsthead
\caption{Continued.}\\
\hline\hline
\multicolumn{1}{c}{\multirow{2}{*}{No}} & \multicolumn{1}{c}{\multirow{2}{*}{Pos}} & \multicolumn{1}{c}{\multirow{2}{*}{$J$}} & \multicolumn{1}{c}{\multirow{2}{*}{\textbf{CV+C+VV~MCDHF}}} & \multicolumn{4}{c}{$\Delta E_{\textbf{(CV+C+VV~MCDHF~(RSMBPT~95\%))-(CV+C+VV~MCDHF)}}$}\\
\cline{5-8}
&&&&\multicolumn{1}{c}{1 var.} & \multicolumn{1}{c}{2 var.} & \multicolumn{1}{c}{3 var.} & \multicolumn{1}{c}{4 var.}  \\
\hline
\noalign{\smallskip}
\endhead
\noalign{\smallskip}
\hline
\hline
\endfoot
\noalign{\smallskip}
 1&  1& 0& -2427.69766& 0.01720& 0.01069& 0.01064& 0.01063 \\
 2&  1& 1& -2427.68992& 0.01376& 0.00967& 0.00965& 0.00965 \\
 3&  1& 2& -2427.67956& 0.01392& 0.00928& 0.00929& 0.00925 \\
 4&  2& 2& -2427.62985& 0.01449& 0.00987& 0.00988& 0.00982 \\
 5&  2& 0& -2427.55745& 0.01781& 0.01153& 0.01143& 0.01143 \\
 6&  2& 1& -2427.01492& 0.01389& 0.00996& 0.00988& 0.00990 \\
 7&  3& 1& -2427.00375& 0.01349& 0.00948& 0.00941& 0.00941 \\
 8&  3& 2& -2427.00215& 0.01374& 0.00942& 0.00938& 0.00938 \\
 9&  3& 0& -2426.99672& 0.01417& 0.00981& 0.00978& 0.00978 \\
10&  4& 1& -2426.98955& 0.01317& 0.00942& 0.00940& 0.00940 \\
11&  1& 3& -2426.98713& 0.01389& 0.01011& 0.01011& 0.01009 \\
12&  4& 2& -2426.98275& 0.01314& 0.00937& 0.00939& 0.00938 \\
13&  5& 1& -2426.97557& 0.01318& 0.00960& 0.00961& 0.00961 \\
14&  5& 2& -2426.96569& 0.01317& 0.00948& 0.00948& 0.00947 \\
15&  4& 0& -2426.93852& 0.01453& 0.01000& 0.00995& 0.00998 \\
16&  2& 3& -2426.83386& 0.01186& 0.00832& 0.00821& 0.00826 \\
17&  3& 3& -2426.83285& 0.01175& 0.00816& 0.00804& 0.00806 \\
18&  6& 2& -2426.83227& 0.01176& 0.00828& 0.00823& 0.00823 \\
19&  1& 4& -2426.83154& 0.01181& 0.00804& 0.00795& 0.00794 \\
20&  4& 3& -2426.81616& 0.01089& 0.00811& 0.00812& 0.00812 \\
21&  2& 4& -2426.81479& 0.01092& 0.00801& 0.00796& 0.00796 \\
22&  1& 5& -2426.81101& 0.01157& 0.00834& 0.00836& 0.00836 \\
23&  5& 3& -2426.81033& 0.01087& 0.00801& 0.00803& 0.00803 \\
24&  7& 2& -2426.81002& 0.01143& 0.00789& 0.00790& 0.00789 \\
25&  8& 2& -2426.80844& 0.01203& 0.00791& 0.00795& 0.00794 \\
26&  6& 1& -2426.80686& 0.01211& 0.00898& 0.00897& 0.00897 \\
27&  3& 4& -2426.80572& 0.01076& 0.00789& 0.00789& 0.00789 \\
28&  7& 1& -2426.80439& 0.01323& 0.00884& 0.00879& 0.00879 \\
29&  9& 2& -2426.80383& 0.01119& 0.00828& 0.00831& 0.00831 \\
30&  8& 1& -2426.80260& 0.01277& 0.00797& 0.00798& 0.00798 \\
31&  5& 0& -2426.80242& 0.01371& 0.00916& 0.00911& 0.00912 \\
32&  9& 1& -2426.80136& 0.01186& 0.00802& 0.00798& 0.00798 \\
33& 10& 2& -2426.79932& 0.01146& 0.00661& 0.00660& 0.00659 \\
34& 11& 2& -2426.79862& 0.01317& 0.00918& 0.00918& 0.00918 \\
35& 10& 1& -2426.79633& 0.01297& 0.00863& 0.00724& 0.00859 \\
36&  6& 3& -2426.79576& 0.01180& 0.00671& 0.00802& 0.00668 \\
37&  6& 0& -2426.79175& 0.01333& 0.00870& 0.00868& 0.00868 \\
38&  4& 4& -2426.79069& 0.01191& 0.00686& 0.00682& 0.00680 \\
39& 11& 1& -2426.78465& 0.01285& 0.00947& 0.00942& 0.00938 \\
40&  2& 5& -2426.78375& 0.01248& 0.00707& 0.00710& 0.00711 \\
41& 12& 2& -2426.78245& 0.01310& 0.00953& 0.00948& 0.00950 \\
42&  7& 3& -2426.78179& 0.01322& 0.00997& 0.00998& 0.00997 \\
43& 12& 1& -2426.77863& 0.01289& 0.00954& 0.00949& 0.00944 \\
44& 13& 2& -2426.77630& 0.01292& 0.00948& 0.00945& 0.00944 \\
45& 13& 1& -2426.77491& 0.01309& 0.00774& 0.00771& 0.00771 \\
46&  7& 0& -2426.77373& 0.01223& 0.00760& 0.00762& 0.00761 \\
47& 14& 1& -2426.77325& 0.01326& 0.00733& 0.00732& 0.00731 \\
48& 14& 2& -2426.77270& 0.01185& 0.00720& 0.00720& 0.00719 \\
49&  8& 3& -2426.77095& 0.01267& 0.00747& 0.00745& 0.00745 \\
50&  5& 4& -2426.76765& 0.01243& 0.00764& 0.00762& 0.00760 \\
51& 15& 2& -2426.76742& 0.01176& 0.00750& 0.00746& 0.00747 \\
52&  8& 0& -2426.76382& 0.01384& 0.00995& 0.00992& 0.00990 \\
53&  9& 3& -2426.75799& 0.01235& 0.00773& 0.00773& 0.00773 \\
54& 16& 2& -2426.75562& 0.01261& 0.00805& 0.00801& 0.00799 \\
55&  9& 0& -2426.73726& 0.01430& 0.00817& 0.00816& 0.00816 \\
56& 17& 2& -2426.73526& 0.01045& 0.00590& 0.00589& 0.00586 \\
57& 15& 1& -2426.73136& 0.01341& 0.00763& 0.00758& 0.00758 \\
58& 10& 3& -2426.72985& 0.01039& 0.00617& 0.00614& 0.00613 \\
59& 11& 3& -2426.72832& 0.01190& 0.00706& 0.00704& 0.00704 \\
60& 18& 2& -2426.72492& 0.01117& 0.00670& 0.00670& 0.00669 \\
61& 16& 1& -2426.72245& 0.01133& 0.00678& 0.00679& 0.00679 \\
62&  6& 4& -2426.72195& 0.01087& 0.00639& 0.00637& 0.00634 \\
63& 19& 2& -2426.72081& 0.01223& 0.00756& 0.00753& 0.00753 \\ \hline
\multicolumn{4}{l}{$\Delta E_{min}$ (in a.u.)} & 0.01039& 0.00590& 0.00589& 0.00586  \\
\multicolumn{4}{l}{$\Delta E_{max}$ (in a.u.)} & 0.01781& 0.01153& 0.01143& 0.01143   \\
\multicolumn{4}{l}{rms (in a.u.)} & 0.01274& 0.00850& 0.00848& 0.00847   \\
\end{longtable}
}

\begin{table*}[!ht]
{\footnotesize
\setlength{\tabcolsep}{3pt}
\caption{The smallest ($\Delta E_{min}$ in a.u.), the largest ($\Delta E_{max}$ in a.u.)
differences and rms of total energies from the MCDHF computations of $OS_1$ and $OS_2$ 
comparing results from regular way and the RSMBPT method.}            
\label{min_max_rms_summary} 
\centering
\begin{tabular}{l r r r r r r r r r}
\hline\hline
&\multicolumn{4}{c}{Even} && \multicolumn{4}{c}{Odd} \\
\cline{2-5} \cline{7-10}
&\multicolumn{1}{c}{1 var.} & \multicolumn{1}{c}{2 var.} & \multicolumn{1}{c}{3 var.} & \multicolumn{1}{c}{4 var.} &&\multicolumn{1}{c}{1 var.} & \multicolumn{1}{c}{2 var.} & \multicolumn{1}{c}{3 var.} & \multicolumn{1}{c}{4 var.}  \\
\hline
\noalign{\smallskip}
\multicolumn{10}{c}{$OS_1$} \\
\multicolumn{10}{c}{Case 95\%} \\
$\Delta E_{min}$ (in a.u.)& 0.01039475& 0.00589933& 0.00588704& 0.00586482 && 0.01078587& 0.00619445& 0.00410314& 0.00378723 \\
$\Delta E_{max}$ (in a.u.)& 0.01781291& 0.01153240& 0.01143447& 0.01143258 && 0.01615159& 0.00844032& 0.00543860& 0.00506363 \\
rms (in a.u.)& 0.01274465& 0.00850122& 0.00847525& 0.00847151 && 0.01345016& 0.00724574& 0.00455746& 0.00430101 \\ \hline
\multicolumn{10}{c}{Case 99\%} \\
$\Delta E_{min}$ (in a.u.)& 0.00479198& 0.00241481& 0.00044516& 0.00044752&& 0.00444358& 0.00147206& 0.00072496& 0.00072521 \\
$\Delta E_{max}$ (in a.u.)& 0.01010324& 0.00488030& 0.00122598& 0.00123373&& 0.00864969& 0.00231023& 0.00107323& 0.00105647 \\ 
rms (in a.u.)& 0.00692783& 0.00366682& 0.00081360& 0.00081380&& 0.00723871& 0.00182232& 0.00086627& 0.00086609 \\ \hline
\multicolumn{10}{c}{Case 99.5\%} \\
$\Delta E_{min}$ (in a.u.)& 0.00403621& 0.00124726& 0.00022377& 0.00022392&& 0.00322708& 0.00075465& 0.00036678& 0.00035865 \\
$\Delta E_{max}$ (in a.u.)& 0.00892285& 0.00265308& 0.00078628& 0.00078657&& 0.00729924& 0.00126755& 0.00055566& 0.00054793 \\
rms (in a.u.)& 0.00591419& 0.00182666& 0.00046031& 0.00045983 && 0.00602911& 0.00094146& 0.00044813& 0.00044605 \\ \hline
\multicolumn{10}{c}{Case 99.95\%} \\
$\Delta E_{min}$ (in a.u.)& 0.00248536& 0.00011430& 0.00002831& 0.00002734&& 0.00352790& 0.00005336& 0.00003057& 0.00003065 \\
$\Delta E_{max}$ (in a.u.)& 0.00591823& 0.00032153& 0.00017859& 0.00017856&& 0.00581758& 0.00013347& 0.00007381& 0.00007382 \\
rms (in a.u.)& 0.00422965& 0.00021375& 0.00007648& 0.00007318&& 0.00479534& 0.00009420& 0.00005156& 0.00005162 \\ \hline
\multicolumn{10}{c}{Case 100\%} \\
$\Delta E_{min}$ (in a.u.)&  0.00003567& 0.00000000& 0.00000000& 0.00000000 && 0.00003871& 0.00000000& 0.00000000& 0.00000000 \\
$\Delta E_{max}$ (in a.u.)&  0.00008260& 0.00000078& 0.00000078& 0.00000078 && 0.00007884& 0.00000001& 0.00000003& 0.00000003\\
rms (in a.u.)&  0.00006240& 0.00000015& 0.00000015& 0.00000015 && 0.00006174& 0.00000001& 0.00000001& 0.00000001 \\ \hline \hline
\multicolumn{10}{c}{$OS_2$} \\
\multicolumn{10}{c}{Case 95\%} \\
$\Delta E_{min}$ (in a.u.)& 0.01514835& 0.00565667& 0.00465128& 0.00476963&& 0.01905315& 0.00664994& 0.00635027& 0.00628709 \\
$\Delta E_{max}$ (in a.u.)& 0.02611409& 0.01186351& 0.00944572& 0.00948070&& 0.02458143& 0.00895267& 0.00824978& 0.00825399 \\
rms (in a.u.)& 0.01959744& 0.00889209& 0.00680218& 0.00676645&& 0.02082576& 0.00747281& 0.00703090& 0.00698012  \\ \hline
\multicolumn{10}{c}{Case 99\%} \\
$\Delta E_{min}$ (in a.u.)& 0.00879110& 0.00085163& 0.00081118& 0.00076419&& 0.01077293& 0.00157237& 0.00150440& 0.00150264 \\
$\Delta E_{max}$ (in a.u.)& 0.01554800& 0.00279595& 0.00235952& 0.00232113&& 0.01502497& 0.00238582& 0.00228452& 0.00228469 \\
rms (in a.u.)& 0.01167577& 0.00205255& 0.00177316& 0.00168195&& 0.01235068& 0.00183188& 0.00177058& 0.00177014 \\ \hline   
\multicolumn{10}{c}{Case 99.5\%} \\
$\Delta E_{min}$ (in a.u.)& 0.00770291& 0.00034064& 0.00041791& 0.00041622&& 0.00862058& 0.00093582& 0.00081631& 0.00081282 \\
$\Delta E_{max}$ (in a.u.)& 0.01361750& 0.00188676& 0.00154804& 0.00153272&& 0.01279172& 0.00150648& 0.00140862& 0.00140949 \\
rms (in a.u.)& 0.00996782& 0.00133000& 0.00107452& 0.00101776&& 0.01058859& 0.00111489& 0.00104843& 0.00104674 \\ \hline 
\multicolumn{10}{c}{Case 99.95\%} \\
$\Delta E_{min}$ (in a.u.)& 0.00388888& 0.00005463& 0.00007514& 0.00007001&& 0.00450417& 0.00014034& 0.00012899& 0.00012840 \\
$\Delta E_{max}$ (in a.u.)& 0.00911764& 0.00047941& 0.00040230& 0.00040681&& 0.00806687& 0.00040195& 0.00038392& 0.00038297 \\
rms (in a.u.)& 0.00571790& 0.00033027& 0.00026441& 0.00026029&& 0.00624877& 0.00024094& 0.00026036& 0.00026023 \\  
\hline
\hline
\end{tabular}
}
\end{table*}

Based on the results of the above investigations, 
it can be concluded that the RSMBPT method can be successfully applied to solve the self-consistent field equations.
Using this method the most important chosen correlations with the specified fraction can be selected, since
to include different types of correlations in the MCDHF calculations in the regular way 
is often a complex task, especially for complex systems.
The radial wavefunctions (obtained using the RSMBPT method) which itself include the most significant correlations can be used
for further RCI calculations.

\subsubsection{Results from RCI calculations}
Tables \ref{min_max_rms_summary_rci} and \ref{en_lev_diff_even2L_95_99.95}
present the results of the RCI computations. 
As it was described above (see Section \ref{Computational_schemes}),
the RCI computations were performed in the regular way and using two different procedures within the framework of RSMBPT method. 
In the first procedure using the RSMBPT method, the radial wavefunctions were taken from the \textbf{CV+C+VV~MCDHF~(RSMBPT)} calculations 
and the CSFs basis was taken from the regular {\sc Grasp}2018 calculations. 
In the second procedure, the radial wavefunctions were taken from the \textbf{CV+C+VV~MCDHF~(RSMBPT)} calculations
and the CSF basis was constructed using the RSMBPT method with the same specified fraction as in the MCDHF.

\begin{table*}[!ht]
{\footnotesize
\setlength{\tabcolsep}{3pt}
\caption{The smallest ($\Delta E_{min}$ in a.u.), the largest ($\Delta E_{max}$ in a.u.)
differences and rms of total energies from the RCI computations of $OS_1$ and $OS_2$ 
comparing results from regular way and using the RSMBPT method.}            
\label{min_max_rms_summary_rci} 
\centering
\begin{tabular}{l r r r r r}
\hline\hline
&\multicolumn{2}{c}{Even} && \multicolumn{2}{c}{Odd} \\
\cline{2-3} \cline{5-6}
&\multicolumn{1}{c}{First procedure} & \multicolumn{1}{c}{Second procedure} &&\multicolumn{1}{c}{First procedure} & \multicolumn{1}{c}{Second procedure}  \\
\hline
\noalign{\smallskip}
\multicolumn{6}{c}{$OS_1$} \\
\multicolumn{6}{c}{Radial wavefunctions from case 95\%} \\
$\Delta E_{min}$ (in a.u.)& 0.00304085& && 0.00000286 \\
$\Delta E_{max}$ (in a.u.)& 0.00769594& && 0.00009646 \\
rms (in a.u.)& 0.00494324& && 0.00004479 \\ \hline
\multicolumn{6}{c}{Radial wavefunctions from case 99\%} \\
$\Delta E_{min}$ (in a.u.)& 0.00000019& && 0.00000016 \\
$\Delta E_{max}$ (in a.u.)& 0.00001724& && 0.00001468 \\ 
rms (in a.u.)& 0.00000680& && 0.00000555 \\ \hline
\multicolumn{6}{c}{Radial wavefunctions from case 99.5\%} \\
$\Delta E_{min}$ (in a.u.)& 0.00000001& && 0.00000034 \\
$\Delta E_{max}$ (in a.u.)& 0.00001541& && 0.00001237 \\
rms (in a.u.)& 0.00000398& && 0.00000637 \\ \hline
\multicolumn{6}{c}{Radial wavefunctions from case 99.95\%} \\
$\Delta E_{min}$ (in a.u.)& 0.00000002& && 0.00000003 \\
$\Delta E_{max}$ (in a.u.)& 0.00000178& && 0.00000189 \\
rms (in a.u.)& 0.00000074& && 0.00000083 \\ \hline
\multicolumn{6}{c}{Radial wavefunctions from case 100\%} \\
$\Delta E_{min}$ (in a.u.)& 0.00000000& && 0.00000000 \\
$\Delta E_{max}$ (in a.u.)& 0.00000006& && 0.00000003 \\
rms (in a.u.)& 0.00000003& && 0.00000001 \\ \hline \hline
\multicolumn{6}{c}{$OS_2$} \\
\multicolumn{6}{c}{Radial wavefunctions from case 95\%} \\
$\Delta E_{min}$ (in a.u.)& 0.00005090& 0.00478032&& 0.00004574& 0.00630025 \\
$\Delta E_{max}$ (in a.u.)& 0.00197267& 0.00952692&& 0.00094435& 0.00826736 \\
rms (in a.u.)& 0.00098455& 0.00678775&& 0.00043676& 0.00698827 \\ \hline
\multicolumn{6}{c}{Radial wavefunctions from case 99\%} \\
$\Delta E_{min}$ (in a.u.)& 0.00000458& 0.00074870&& 0.00000710& 0.00151110 \\
$\Delta E_{max}$ (in a.u.)& 0.00023362& 0.00232700&& 0.00029347& 0.00229393 \\
rms (in a.u.)& 0.00014306& 0.00167533&& 0.00009579& 0.00177984 \\ \hline   
\multicolumn{6}{c}{Radial wavefunctions from case 99.5\%} \\
$\Delta E_{min}$ (in a.u.)& 0.00000024& 0.00041214&& 0.00000244& 0.00081798 \\
$\Delta E_{max}$ (in a.u.)& 0.00008720& 0.00154291&& 0.00014983& 0.00141603 \\
rms (in a.u.)& 0.00004940& 0.00102058&& 0.00004814& 0.00105443 \\ \hline 
\multicolumn{6}{c}{Radial wavefunctions from case 99.95\%} \\
$\Delta E_{min}$ (in a.u.)& 0.00000018& 0.00007105&& 0.00000004& 0.00012925 \\
$\Delta E_{max}$ (in a.u.)& 0.00002741& 0.00041042&& 0.00001590& 0.00038453 \\
rms (in a.u.)& 0.00000952& 0.00026222&& 0.00000824& 0.00026121 \\  
\hline
\hline
\end{tabular}
}
\end{table*}

Table \ref{min_max_rms_summary_rci} presents the summary of the rms, of the smallest ($\Delta E_{min}$) and the largest ($\Delta E_{max}$)
differences comparing the total energies from the \textbf{CV+C+VV~RCI~(RSMBPT)} method with the regular 
{\sc Grasp}2018 calculations (\textbf{CV+C+VV~RCI}).
As seen from the Table, the results (these in columns 2 and 4) from the first procedure using the RSMBPT method
in the case of ‘100\%’ reproduce the results from regular calculations.
Using the radial wavefunctions from the \textbf{CV+C+VV~MCDHF~(RSMBPT)} calculations 
with a decreased amount of these correlations (99.95, 99.5, 99, 95\%) the differences
increase, but the results are still in good agreement with regular calculations.

In the second procedure, when the RSMBPT method is applied to obtain the radial wavefunctions and the CSF basis (results in columns 3 and 5), 
the differences between the two calculations are even larger.
The rms in the case, when 95\% of the CV, C, VV correlations are included in both (MCDHF and RCI) computations, is equal 0.00678775 a.u. 
and $\Delta E_{max}$ for this computation is equal 0.00952692 a.u. (or 0.00039276\%).
By increasing the amount of these correlations the differences decreases, and in the case, when 99.95\% 
of the CV, C, VV correlations are included in both (MCDHF and RCI) computation, the rms is equal 0.00026222 a.u. 
and $\Delta E_{max}$ for this computation is equal 0.00041042 a.u. (or 0.00001693\%).

Table \ref{en_lev_diff_even2L_95_99.95} shows the comparison of energy levels from
regular {\sc Grasp}2018 calculations and calculations using the RSMBPT method for $OS_2$ even states.
In the Table energy levels from regular calculations and differences 
between the \textbf{CV+C+VV~RCI~(RSMBPT)} and \textbf{CV+C+VV~RCI} calculations
are given. The results of the \textbf{CV+C+VV~RCI~(RSMBPT)} calculations are given only 
for two boundary cases (when radial wavefunctions are taken from the \textbf{CV+C+VV~MCDHF~(RSMBPT)} calculations 
in two cases 99.95 and 95\%).
The rms deviations obtained for the first procedure RCI calculations from the regular {\sc Grasp}2018 data
are 276.96 cm$^{-1}$ and 2.32 cm$^{-1}$, respectively, when 
95 or 99.95\% of the CV, C, VV correlations are included in the MCDHF.
It should be noted that energies of the computed states are up to 216000 cm$^{-1}$.
The differences increase when the RSMBPT method is used for MCDHF and RCI computations (the second procedure).
In this procedure, the rms deviations from the regular {\sc Grasp}2018 data
are 509.08 cm$^{-1}$ and 25.92 cm$^{-1}$, respectively, when 
95 or 99.95\% of the CV, C, VV correlations are included in both (MCDHF and RCI) computations.
The largest difference between regular calculations and calculations when 95\% of the CV, C, VV correlations 
are included in both (MCDHF and RCI) computations is equal 888.99 cm$^{-1}$ (or 0.421\%), and in case 99.95\% it is equal 49.73 cm$^{-1}$ (or 0.025\%).

{\footnotesize
\begin{longtable}{r r r r r r r r r}
\caption{\label{en_lev_diff_even2L_95_99.95} Energy levels (in cm$^{-1}$) from \textbf{CV+C+VV~RCI} calculations and differences (in cm$^{-1}$) 
between \textbf{CV+C+VV~RCI~(RSMBPT)} and \textbf{CV+C+VV~RCI} energies ($\Delta E_{\textbf{(CV+C+VV~RCI~(RSMBPT))-(CV+C+VV~RCI)}}$) 
for $OS_2$ even states of Se~III are given when CV, C and VV correlations are included in the computations.}\\
\hline\hline
\multicolumn{1}{c}{\multirow{3}{*}{No}} & \multicolumn{1}{c}{\multirow{3}{*}{Pos}} & \multicolumn{1}{c}{\multirow{3}{*}{$J$}} & \multicolumn{1}{c}{\multirow{3}{*}{\textbf{CV+C+VV~RCI}}} & \multicolumn{5}{c}{$\Delta E_{\textbf{(CV+C+VV~RCI~(RSMBPT))-(CV+C+VV~RCI)}}$}\\
\cline{5-9} 
&&&& \multicolumn{2}{c}{First procedure} && \multicolumn{2}{c}{Second procedure}\\
\cline{5-6} \cline{8-9}
&&&&\multicolumn{1}{c}{95\%} & \multicolumn{1}{c}{99.95\%} && \multicolumn{1}{c}{95\%} & \multicolumn{1}{c}{99.95\%}  \\
\hline
\noalign{\smallskip}
\endfirsthead
\caption{Continued.}\\
\hline\hline
\multicolumn{1}{c}{\multirow{3}{*}{No}} & \multicolumn{1}{c}{\multirow{3}{*}{Pos}} & \multicolumn{1}{c}{\multirow{3}{*}{$J$}} & \multicolumn{1}{c}{\multirow{3}{*}{\textbf{CV+C+VV~RCI}}} & \multicolumn{5}{c}{$\Delta E_{\textbf{(CV+C+VV~RCI~(RSMBPT))-(CV+C+VV~RCI)}}$}\\
\cline{5-9} 
&&&& \multicolumn{2}{c}{First procedure} && \multicolumn{2}{c}{Second procedure}\\
\cline{5-6} \cline{8-9}
&&&&\multicolumn{1}{c}{95\%} & \multicolumn{1}{c}{99.95\%} && \multicolumn{1}{c}{95\%} & \multicolumn{1}{c}{99.95\%}  \\
\hline
\noalign{\smallskip}
\endhead
\noalign{\smallskip}
\hline
\hline
\endfoot
\noalign{\smallskip}
 1& 1	&0&	0.00     & 0.00   & 0.00  && 0.00   & 0.00     \\
 2& 1	&1&	1682.62  & 3.75   & -0.32 && -330.09& -21.68   \\
 3& 1	&2&	3889.34  & 6.04   & -0.88 && -411.08& -24.74   \\
 4& 2	&2&	13797.80 & 23.17  & 1.86  && -312.40& -20.79   \\
 5& 2	&0&	29482.86 & 82.02  & 2.96  && 152.76 & -4.49    \\
 6& 2	&1&	151160.29& -98.23 & -3.90 && -257.70& 27.52    \\
 7& 3	&1&	153592.00& -99.55 & -2.97 && -330.38& 25.41    \\
 8& 3	&2&	153914.82& -102.93& -3.31 && -325.03& 25.14    \\
 9& 3	&0&	155152.31& -89.78 & -1.29 && -193.47& 34.01    \\
10& 4	&1&	156676.51& -91.53 & -2.96 && -341.68& 26.93    \\
11& 1	&3&	157184.53& -103.16& -4.01 && -140.65& 35.61    \\
12& 4	&2&	158137.34& -89.27 & -3.00 && -322.07& 27.12    \\
13& 5	&1&	159722.61& -95.52 & -3.69 && -354.90& 26.77    \\
14& 5	&2&	161643.32& -86.43 & -2.54 && -314.04& 26.27    \\
15& 4	&0&	167467.38& -75.92 & -3.37 && -122.41& 27.77    \\
16& 2	&3&	191065.56& -528.12& -0.05 && -485.28& 43.60    \\
17& 3	&3&	191311.15& -529.97& 1.45  && -514.99& 41.16    \\
18& 6	&2&	191438.47& -548.76& -0.15 && -507.27& 45.48    \\
19& 1	&4&	191591.48& -539.72& 1.80  && -517.79& 40.19    \\
20& 4	&3&	194890.31& -530.86& -1.54 && -551.92& 41.63    \\
21& 7	&2&	195071.18& -165.89& 0.05  && -521.09& -10.96   \\
22& 2	&4&	195200.75& -530.91& -0.13 && -546.13& 39.68    \\
23& 1	&5&	196006.00& -515.51& 0.22  && -393.95& 49.73    \\
24& 5	&3&	196208.59& -536.94& 0.38  && -561.74& 40.04    \\
25& 8	&2&	196466.68& -548.38& -1.25 && -570.10& 43.26    \\
26& 6	&1&	196813.18& -252.84& -0.55 && -503.12& 0.60     \\
27& 7	&1&	197037.20& -284.25& -2.11 && -540.48& 35.23    \\
28& 3	&4&	197214.70& -598.63& -0.45 && -518.21& 47.03    \\
29& 8	&1&	197246.38& -290.55& -0.44 && -544.53& -10.35   \\
30& 5	&0&	197413.53& -119.00& -0.72 && -335.69& 7.73     \\
31& 9	&2&	197572.23& -505.66& 0.61  && -759.66& 24.24    \\
32& 10	&2&	197711.31& -314.91& 0.10  && -526.22& 4.17    \\
33& 9	&1&	197869.14& -249.40& -2.31 && -296.61& 40.00    \\
34& 6	&3&	198439.99& -280.51& 0.28  && -741.17& -12.80   \\
35& 10	&1&	198691.21& -122.59& -1.98 && -323.69& 18.69   \\
36& 11	&2&	198745.26& -161.50& -2.13 && -319.87& 29.04   \\
37& 6	&0&	199454.22& -118.96& -1.28 && -229.36& 24.29    \\
38& 4	&4&	199536.16& -265.27& 0.51  && -655.89& -8.93    \\
39& 2	&5&	201028.11& -254.34& 0.55  && -510.49& 5.09     \\
40& 11	&1&	201813.06& -100.54& -4.12 && -318.12& 31.11   \\
41& 12	&2&	202295.91& -102.62& -4.02 && -253.44& 32.55   \\
42& 7	&3&	202470.62& -103.20& -3.94 && -152.60& 34.19    \\
43& 12	&1&	203148.82& -105.64& -3.87 && -546.49& 30.00   \\
44& 7	&0&	203351.47& -215.51& -0.54 && -610.10& -12.55   \\
45& 13	&1&	203404.08& -206.94& -0.46 && -552.38& -15.50  \\
46& 13	&2&	203527.76& -181.36& -1.88 && -766.05& 1.90    \\
47& 14	&2&	203618.74& -137.47& -1.61 && -322.28& 10.36   \\
48& 8	&3&	203950.72& -203.44& 0.25  && -665.57& -12.78   \\
49& 14	&1&	204211.81& -110.85& -0.82 && -593.82& -16.51  \\
50& 5	&4&	204652.02& -190.72& 0.44  && -547.98& -6.71    \\
51& 15	&2&	205754.78& -110.57& -1.17 && -678.00& -16.14  \\
52& 8	&0&	206137.41& -88.19 & -4.35 && -133.73& 29.68    \\
53& 16	&2&	206141.26& -139.46& 1.66  && -386.88& 1.47    \\
54& 9	&3&	207759.36& -115.37& -1.37 && -591.68& -11.84   \\
55& 17	&2&	211050.37& -339.75& 5.05  && -888.99& 0.95    \\
56& 9	&0&	212148.21& -122.03& -4.03 && -457.05& -12.15   \\
57& 10	&3&	212253.84& -336.77& 3.95  && -783.63& 7.51    \\
58& 11	&3&	213384.11& -189.00& 1.09  && -690.00& -14.25  \\
59& 15	&1&	213402.16& -123.00& -3.91 && -636.39& -14.85  \\
60& 6	&4&	213959.74& -332.19& 2.60  && -651.72& 11.21    \\
61& 18	&2&	214098.17& -219.75& -0.92 && -870.78& -19.97  \\
62& 16	&1&	214627.71& -237.65& -2.08 && -854.12& -20.46  \\
63& 19	&2&	215656.63& -125.53& -3.35 && -658.07& -14.78  \\ \hline
\multicolumn{4}{l}{$N_{CSFs}$}        & 2923523& 2923523&& 1148711& 2567016 \\
\multicolumn{4}{l}{rms (in cm$^{-1}$)}& 276.96& 2.32&& 509.08& 25.92   \\
\end{longtable}                                                     
}

\section{Conclusions}
\label{Sec:Conclusions}

The method, based on the Rayleigh-Schr\"odinger perturbation theory in an irreducible tensorial form, 
is extended to estimate the valence-valence correlations, which are described 
by the three-particle Feynman diagram. 
The expressions to calculate the influence of these correlations are provided,
also additional developments to calculate the spin-angular parts of three-particle Feynman diagram has been made 
to the program library \texttt{librang} of the {\sc Grasp}.
This extended RSMBPT method allows to estimate the contribution 
of any $K'$ configuration of the CV, C, CC and VV correlations 
with preferred core, virtual orbitals sets and with any number of valence electrons for any atom or ion.

This is the first time that CSF bases constructed using the RSMBPT method 
have been used to solve the self-consistent field equations.
Previously, this method was only applied to RCI computations. Thus, this work demonstrates a third way of the application 
of the RSMBPT method in atomic calculations (other two ways of it application 
are presented in a series of previous papers by G. Gaigalas, P. Rynkun and L. Kitovienė).

The use of the RSMBPT method in the MCDHF computations 
enables the inclusion of the most significant correlations of various types, 
which is not very feasible in regular {\sc Grasp} calculations, especially for complex systems.
Based on the results obtained in this study work, it can be concluded that the RSMBPT method can be successfully applied 
to solve the self-consistent field equations.

The radial wavefunctions obtained using the RSMBPT method
can be used in the RCI computations in several procedures: i) these can be used with the extended CSF basis; or
ii) these can be used with CSF basis constructed using the RSMBPT method with the specified fraction as in the MCDHF.
Results of both RCI computations are in good agreement with the results of the regular RCI computations.


\end{document}